\documentclass[preprintnumbers,superscriptaddress,amsmath,amssymb,twocolumn]{revtex4-2}
\usepackage{float}
\usepackage{graphicx}
\usepackage{dcolumn}
\usepackage{bm}
\usepackage{epstopdf}
\usepackage{xcolor}
\usepackage{textcomp}
\usepackage{soul}
\usepackage{csquotes}
\usepackage{amsbsy}
\usepackage{amssymb}
\usepackage{mathrsfs} 
\usepackage{amsmath}
\usepackage{bigints} 
\usepackage{amsthm,amssymb} 
\usepackage[utf8]{inputenc}
\usepackage[shortlabels]{enumitem}
\usepackage{float}
\usepackage{mathrsfs,bigints,mathtools,dsfont}
\usepackage[colorlinks=true,linkcolor=blue,citecolor=blue]{hyperref}%
\usepackage[toc]{appendix}
\begin{document}
\title{Synchronization of topological signals in higher-order adaptive multilayer network}
\author{Palash Kumar Pal}\email{palash93@gmail.com}\affiliation{Physics and Applied Mathematics Unit, Indian Statistical Institute, 203 B. T. Road, Kolkata 700108, India}
\author{Dibakar Ghosh}\email{dibakar@isical.ac.in}
\affiliation{Physics and Applied Mathematics Unit, Indian Statistical Institute, 203 B. T. Road, Kolkata 700108, India}	
\author{J\"urgen Kurths}\email{kurths@pik-potsdam.de}
\affiliation{Potsdam Institute for Climate Impact Research, Potsdam, Germany}
\affiliation{Department of Physics, Humboldt-Universit\"at zu Berlin, Berlin, Germany}
\begin{abstract}
    The study of synchronization in complex systems has recently been revolutionized by incorporating higher-order interactions through simplicial complexes. Building in particular upon the higher-order Kuramoto model, which considers oscillators on nodes, links, and higher-dimensional simplices. This work extends the monolayer framework of the higher-order Kuramoto model to multilayer networks where the layers are adaptively coupled through order parameters of the oscillators placed on the simplices. We propose two multilayer architectures: one that allows interactions between signals of the same dimension across layers and the other that permits cross-dimensional interactions. We observe that a higher coupling strength is required for synchronization transitions of the node signals and the projected uplink and downlink signals during adaptation. For example, incorporating node dynamics into link evolution delays the onset of synchronization. This study opens an avenue for understanding complex dynamical processes within interconnected higher-order structures. Finally, we present a comprehensive theoretical framework, first for a bilayer network where layers are random networks treated under the annealed approximation, and then extend the analysis to the case of fully connected layers. The theoretical predictions align remarkably well with numerical simulations, accurately capturing the dynamics of the original model in a globally coupled scenario.
\end{abstract}
\maketitle
\section{Introduction}
\par The study of synchronization phenomena has long been a central topic in the analysis of coupled complex systems \cite{synchronization1,synchronization2}, with far-reaching implications in fields ranging from neuroscience \cite{abarbanel1996synchronisation,cao2006adaptive} to power grids \cite{nishikawa2015comparative} and social dynamics \cite{morales2017global,dikker2017brain}. Since its inception, the Kuramoto model has served as a paradigmatic framework for understanding synchronization through pairwise interactions between phase oscillators \cite{acebron2005kuramoto}. However, many real-world systems, ranging from the brain to social systems, exhibit interactions that are intrinsically multi-directional involving groups rather than pairs of units \cite{battiston2021physics,majhi2022dynamics,majhi2024patterns,gao2023dynamics,zhang2023higher}. This observation has led to the development of higher-order network models, such as simplicial complexes and hypergraphs, which capture interactions not only between pairs of elements but also between triplets, quadruplets, and higher-dimensional groupings through different-dimensional simplices and hyperedges included, respectively, in the simplicial complex, hypergraphs \cite{bianconi2021higher}. By extending pairwise coupling to multi-body interactions, higher-order Kuramoto models reveal fundamentally distinct synchronization mechanisms. Such interactions can restructure the collective phase space, producing abrupt synchronization transitions, multistability, and dynamical regimes that have no analog in traditional pairwise-coupled oscillator networks \cite{tanaka2011multistable,skardal2020higher,skardal2021higher,tang2022optimizing,bick2022multi}. \par Yet, these approaches to synchronization of phase oscillators of the Kuramoto model have focused on pairwise interactions and higher-order interactions among oscillators placed on the nodes of a network \cite{acebron2005kuramoto,rodrigues2016kuramoto}.

\par Recent advances have extended dynamical models to simplicial complexes, where topological signals are defined on higher-dimensional simplices such as links, triangles, and beyond \cite{millan2020explosive,barbarossa2020topological,ghorbanchian2021higher,millan2022geometry,giambagli2022diffusion,arnaudon2022connecting,torres2020simplicial,ziegler2022balanced,carletti2023global,calmon2023local,calmon2023dirac,gong2024higher,muolo2024three,nurisso2024unified,millan2025topology,zaid2026designing}. The higher-order Kuramoto model introduced by Mill\'an et al. \cite{millan2020explosive} has overcome the limitation that the dynamics must be confined to the nodes of a simplicial complex. They demonstrated that when solenoidal and irrotational components of link signals are adaptively coupled, an explosive synchronization transition can emerge, marking a significant departure from the smooth, continuous transitions characteristic of the Kuramoto model. Following the development of the higher-order Kuramoto model, research on explosive simplicial synchronization \cite{ghorbanchian2021higher}, global topological synchronization \cite{carletti2023global,wang2024global,wang2026topology} has gained significant attention, and also $D$-dimensional topological Kuramoto model has emerged by Wang et al. \cite{wang2025higher}. Additionally, Dirac synchronization has further illustrated how the coupling between node and edge signals through the topological Dirac operator can lead to rich dynamical behaviors, including hysteresis and rhythmic phases, on fully connected networks \cite{bianconi2021topological,calmon2022dirac,calmon2023local,carletti2025global}.

\par Parallel to these developments, the framework of topological signal processing has provided essential tools for understanding signals defined on higher-order structures \cite{barbarossa2020topological}. The topological signal processing generalizes graph signal processing by addressing signals not only on nodes but also on edges and higher simplices, with applications ranging from flow analysis in communication networks to the inference of biological regulatory interactions. Moreover, recent work on multilayer simplicial complexes has highlighted their crucial role in capturing the interdependence of multiple interacting layers, offering a more realistic framework for modeling and understanding the collective behavior of complex systems \cite{majhi2016chimera,majhi2019chimera,anwar2022stability,anwar2022intralayer,pal2023desynchrony,pal2024global,li2025synchronization}. Despite significant progress in extending the Kuramoto model to higher-order and multilayer networks, a fundamental gap remains. Existing studies of higher-order multilayer networks typically confine the dynamics to node-level phase variables \cite{kachhvah2021explosive,jalan2022multiple,rathore2023synchronization,ghosh2025transitions}, thereby neglecting the inherently higher-dimensional interactions encoded in the simplicial structure of each layer. Conversely, higher-order Kuramoto models that assign phase variables to simplices of all dimensions have so far been restricted to single-layer networks \cite{ghorbanchian2021higher,millan2020explosive}, preventing the exploration of interlayer coupling effects on higher-order dynamics.

\par Therefore, a unified framework capable of describing the synchronization of topological signals—namely, dynamical phase variables defined on simplices of different dimensions across multiple layers—is still lacking. Such a framework is essential for capturing the interplay between higher-order interactions and multilayer organization, which naturally arises in many real-world systems. To address this open problem, in this paper we develop and analyze a Kuramoto-type model on a multilayer network of simplicial complexes and investigate the resulting synchronization phenomena. Specifically, we consider bilayer systems in which each layer is represented by a two-dimensional simplicial complex and supports dynamical phase variables on nodes and links, together with projections of link phases onto adjacent simplicial spaces. The two layers are coupled through adaptive interlayer interactions that may act either between signals of the same topological dimension or across different dimensions. This construction allows us to systematically disentangle the roles of node-, link-, and triangle-related signals and to assess how cross-dimensional coupling reshapes synchronization transitions. By combining numerical simulations with analytical results obtained in the annealed and fully connected limits, we show that adaptive multiplex coupling can significantly modify the onset of synchronization and, in particular, induce robust delays in the synchronization of higher-order signals. 
\section{Results and Discussion}
The higher-order Kuramoto model extends the classical Kuramoto framework by describing the collective dynamics of oscillators defined not only on nodes but also on higher-dimensional simplices, such as links, filled triangles, and higher-order structures. This formulation enables the study of synchronization of topological signals, namely phases residing on simplices of different dimensions, and the interactions among them. \\ So far, synchronization of topological signals has been investigated almost exclusively within a single-layer framework. In this context, Millán et al. \cite{millan2020explosive} showed that the synchronization transitions of node phases and the projections of link phases onto adjacent simplices (nodes and triangles) are continuous when these projected phases are uncoupled. In contrast, when the two projected phases are coupled, the system exhibits explosive synchronization, characterized by a discontinuous transition. Subsequently, Ghorbanchian et al. \cite{ghorbanchian2021higher} extended this framework by incorporating additional couplings between node phases and down-link phases, considering both coupled and uncoupled uplink–downlink configurations. They demonstrated that such couplings can induce explosive topological synchronization, in which node and link phases synchronize simultaneously via a discontinuous transition. \\ Despite these advances, the role of inter-layer interactions in shaping topological synchronization remains largely unexplored. In particular, how topological signals defined on simplices of different dimensions synchronize when embedded in a multilayer structure is still an open question.
\begin{figure*}[t]
    \centering
    \includegraphics[width=1.0\textwidth]{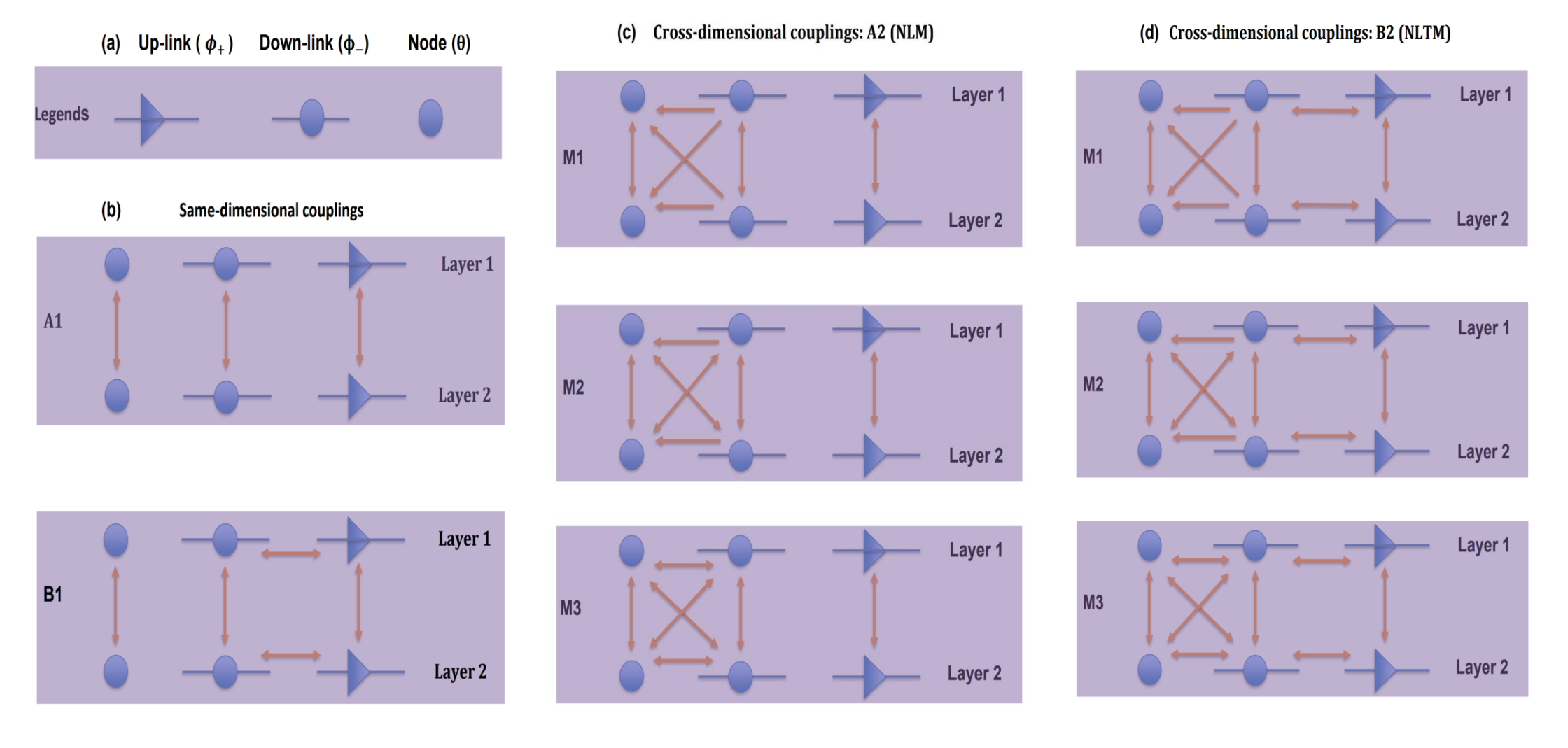}
    \caption{\textbf{Schematic diagram of coupling between two layers of the multilayer network.} Arrowed lines represent couplings between signals (phases); an arrow pointing toward a signal indicates that this signal is coupled by the signal from which the arrow originates. (a) Symbols representing the phases: node phase (filled circle), projection of a link phase onto a node (line with a filled circle), and projection of a link phase onto a triangle (line with a filled triangle). (b) Models A1 and A2 with same-dimensional coupling. (c) Cross-dimensionally coupled models M1, M2, and M3 of model A2 (node–link multilayer, NLM). (d) The corresponding three cross-dimensional couplings of model B2 (node–link–triangle multilayer, NLTM).}
    \label{schematic_diag}
\end{figure*}
\\ Motivated by this gap, we extend the higher-order Kuramoto model to a multilayer setting, where layers are coupled through multiple same-dimensional and cross-dimensional interaction schemes, as illustrated in the schematic diagram, Fig. \ref{schematic_diag}. Figure~\ref{schematic_diag} provides a schematic summary of all multilayer coupling configurations explored in this work. Each layer consists of a two-dimensional simplicial complex supporting node and link phase dynamics, while triangles are involved through projections of link phases. 
In the same-dimensional setting, Fig. \ref{schematic_diag}(b), the two layers are coupled only through signals of the same topological dimension. In this setting, there are two models, A1 and B1. In contrast, the cross-dimensional models introduce adaptive couplings between signals of different dimensions across layers. Figure \ref{schematic_diag}(c) represents node-link multilayer (NLM) configurations in which link dynamics in one layer are adaptively coupled to node or projected link signals in the other layer. 
Figure~\ref{schematic_diag}(d) extends this construction to node--link--triangle multilayers (NLTM), where link dynamics is additionally coupled to triangle-level projections. 
This schematic highlights how progressively richer cross-dimensional interactions are introduced, allowing us to disentangle the role of node-, link-, and triangle-level signals in shaping synchronization transitions. In these two settings, there are three models, M1, M2, and M3, which will be defined later. This framework allows us to systematically investigate how inter-layer couplings influence the synchronization transitions of topological signals across different simplices and layers.
\\We introduce a higher-order multilayer topological Kuramoto model to investigate the synchronization of topological signals defined on simplices and their projections onto consecutive upper and lower dimensional simplices within each layer of the multilayer network. Specifically, we consider a bilayer model of a multilayer network constituted by two simplicial complexes of dimension two, denoted as $SC_1$ and $SC_2$ (see "Methods"). The simplicial complexes, $SC_1$ and $SC_2$ consist of $N^{(l)}_d$ ($l=1,2$) number of simplices of dimension $d$, such as $N^{(l)}_0$ nodes ($0$-simplices), $N^{(l)}_1$ links ($1$-simplices), $N^{(l)}_2$ filled triangles ($2$-simplices) in the simplicial complex $SC_l$, subject to the condition $N^{(1)}_0 = N^{(2)}_0$, ensuring that both layers share the same set of nodes.

To formulate the simplicial dynamics, we employ tools from algebraic topology. In particular, we denote the $n$-th incidence matrix in the layer $l$ as $B_n^{(l)}$, corresponding to the discrete boundary operator of order $n$ for that specific layer (see ``Methods"). These incidence matrices encode the adjacency relations between simplices of consecutive dimensions and provide the structural foundation for defining higher-order coupling.

Now, let the topological signals of the nodes and links in the layer-$l$ be given by the co-chains defined by the column vectors ${\Theta}^{(l)}=(\theta^{(l)}_1,\theta^{(l)}_2,\dots,\theta^{(l)}_{N^{(l)}_0})^T$, and  ${\Phi}^{(l)}=(\phi^{(l)}_1,\phi^{(l)}_2,\dots,\phi^{(l)}_{N^{(l)}_1})^T$, respectively ($T$ means transpose of matrix). The projections of the link signals onto the node space and the triangle space within each layer $l$ through the incidence matrices $B^{(l)}_1$ and $B^{(l)}_2$ are defined, respectively, as
\begin{equation}
  \begin{split}
    {\Phi}^{(l)}_-=B^{(l)}_1{\Phi}^{(l)},\\{\Phi}^{(l)}_+=[B^{(l)}_2]^T{\Phi}^{(l)}. 
  \end{split} 
  \label{proj}
\end{equation} 
Before going on to formulate the dynamics of the topological signals, let us introduce all the order parameters required to model our multilayer framework.
\par The order parameters for two-layer are as follows:\\
For the node and link signals of the two layers are, respectively,
\begin{equation}
  R^{(l)}_0 = \left| \frac{1}{N_0^{(l)}} \sum_{j=1}^{N_0^{(l)}} e^{\mathbf{i} \theta^{(l)}_j} \right|,
  R^{(l)} = \left| \frac{1}{N_1^{(l)}} \sum_{j=1}^{N_1^{(l)}} e^{\mathbf{i} \phi^{(l)}_j} \right|,  
\end{equation}
where $\mathbf{i}=\sqrt{-1}$.
 \\ For the projections of the link signals onto the triangles and nodes, given in Eq. \eqref{proj}, within the two layers are, respectively,
\begin{equation}
  R^{(l)}_+ = \left| \frac{1}{N_2^{(l)}} \sum_{j=1}^{N_2^{(l)}} e^{\mathbf{i} \phi^{(l)}_{+j}} \right|,R^{(l)}_- = \left| \frac{1}{N_0^{(l)}} \sum_{j=1}^{N_0^{(l)}} e^{\mathbf{i} \phi^{(l)}_{-j}} \right|.  
\end{equation}
\\ The amplitudes of the order parameters sensitive to the solenoidal and irrotational components of the two layers link signals are, respectively,
\begin{equation}
 R^{(l)}_1 = \left| \frac{1}{N_1^{(l)}} \sum_{j=1}^{N_1^{(l)}} e^{\mathbf{i} y^{(l)}_{1j}} \right|,R^{(l)}_2 = \left| \frac{1}{N_1^{(l)}} \sum_{j=1}^{N_1^{(l)}} e^{\mathbf{i} y^{(l)}_{2j}} \right|.   
\end{equation}
Here, $y_{ij}^{(l)}$ are the components of the vectors $Y^{(l)}_1$ and $Y^{(l)}_2$ given in the Methods Section \ref{hodge}.
\par The classic Kuramoto models of the isolated layers, i.e., \begin{equation}
\dot{\theta}_i^{(l)}=\omega_i^{(l)}+\sigma\sum_{j=1}^{N_0^{(l)}}a_{i,j}^{(l)}\sin{(\theta_j^{(l)}-\theta_i^{(l)})},
\end{equation} can be written in terms of the boundary matrix as, 
\begin{equation}
    \dot{\Theta}^{(l)}={\Omega}^{(l)}-\sigma B^{(l)}_1\sin{([B^{(l)}_1]^T{\Theta}^{(l)})},
\end{equation} where $a_{i,j}^{(l)}$'s are the adjacency matrix elements of the network skeletons of the simplicial complexes in the $l$-th layer, ${\Omega}^{(l)}=(\omega^{(l)}_1,\omega^{(l)}_2,\dots,\omega^{(l)}_{N^{(l)}_0})^T$ is the natural frequency vector of the node signal, and $\sigma$ is the coupling strength.  
\\Likewise, the evolution of the link signals of the isolated layers can be described in terms of the boundary matrices as,
\begin{equation}
 \dot{\Phi}^{(l)}={\tilde{\Omega}}^{(l)}-\sigma[B^{(l)}_1]^T\sin{(B^{(l)}_1{\Phi}^{(l)})}-\sigma B^{(l)}_2\sin{([B^{(l)}_2]^T{\Phi}^{(l)})},  
\end{equation} where ${\tilde{\Omega}}^{(l)}$ is the natural frequency vector of the link signal.
\\However, the dynamics of the two projected phases ${\Phi}^{(l)}_-$ and ${\Phi}^{(l)}_+$ will follow two differential equations that are not coupled. Millan et al. \cite{millan2020explosive} introduced an adaptive coupling by which these two projected phases can be coupled. The evolution equation for this adaptive coupling is 
\begin{equation}
\begin{split}
 \dot{\Phi}^{(l)}={\tilde{\Omega}}^{(l)}-\sigma R^{(l)}_+[B^{(l)}_1]^T\sin{(B^{(l)}_1{\Phi}^{(l)})}\\-\sigma R^{(l)}_-B^{(l)}_2\sin{([B^{(l)}_2]^T{\Phi}^{(l)})}. 
 \end{split}
\end{equation}
Till now, we have not coupled the layers yet. Therefore, to create a multilayer model, we now couple the layers through the amplitude of the order parameters of the phases in the layers. 

\par There exist several options to create an adaptive multilayer model. Let's start to create a multilayer network by following a particular approach where only the interaction between signals of the same dimension across the layers is allowed.\\\\
\subsection{Multilayer coupling between same-dimensional signals} In this model, the node, up-link, down-link signals of one layer can modulate the strength of the node, up-link, down-link signals, respectively, of the other layer.
The differential equation that describes the evolution of the nodes is given as
\begin{equation}
    \dot{\Theta}^{(l)} = \Omega^{(l)} - \sigma R_0^{(l')}B^{(l)}_1 \sin\left([B^{(l)}_1]^T{ \Theta}^{(l)}\right),~ l'\neq l\in\{1,2\}.
    \label{eq_sd_node}
\end{equation}
and the links in the multilayer network can evolve according to two distinct mechanisms. Based on these evolution rules, we consider two classes of multilayer frameworks, denoted by A1 and B1: \begin{enumerate}
\renewcommand{\labelenumi}{\Alph{enumi}1.}
    \item Projections on nodes and triangles in respective layers are not coupled; the dynamical equation is given by
\begin{equation}
    \begin{split}
\dot{\Phi}^{(l)} = \tilde{\Omega}^{(l)} - \sigma R^{(l')}_- [B^{(l)}_1]^T \sin\left({\Phi}^{(l)}_-\right)- \sigma R^{(l')}_+ [B^{(l)}_2] \sin\left({\Phi}^{(l)}_+\right).   
\end{split}
\label{eq_sd_link1}
\end{equation}
\item The projections within their layers are coupled, and the evolution becomes
\begin{equation}
    \begin{split}
\dot{\Phi}^{(l)} = \tilde{\Omega}^{(l)} - \sigma R^{(l)}_+ R^{(l')}_- [B^{(l)}_1]^T \sin\left({\Phi}^{(l)}_-\right)\\- \sigma R^{(l)}_- R^{(l')}_+ [B^{(l)}_2] \sin\left({ \Phi}^{(l)}_+\right).
\end{split}
\label{eq_sd_link2}
\end{equation}
\end{enumerate}

\begin{figure}
    \centering
    \includegraphics[width=1\linewidth]{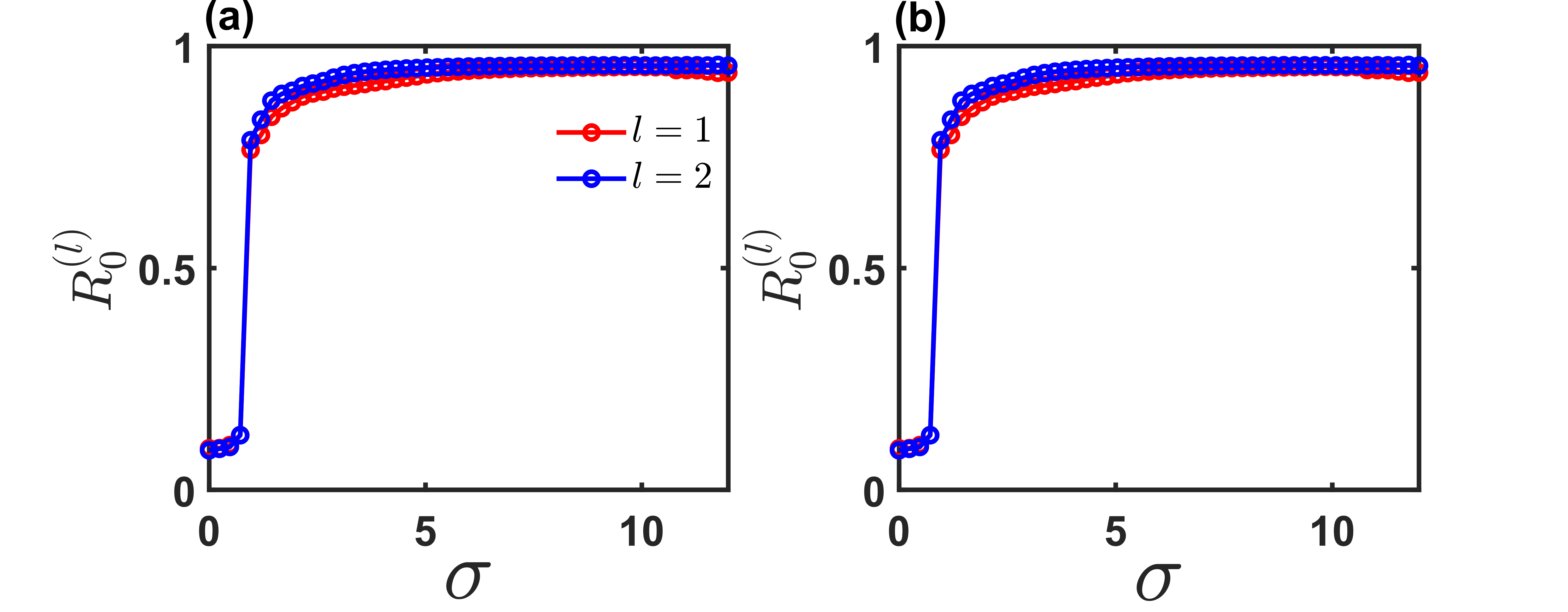}
    \caption{\textbf{Synchronization of the nodes of the layers in the multilayer network model when the same-dimensional signals are interacting and different-dimensional signals are not.} The amplitudes of the order parameters for nodes when the projections of the links on the adjacent dimensional signals are (a) not coupled in the link dynamics (given by A1) and (b) when they are coupled (given by B1).}
    \label{fig_a}
\end{figure}
\begin{figure}
    \centering
    \includegraphics[width=1\linewidth]{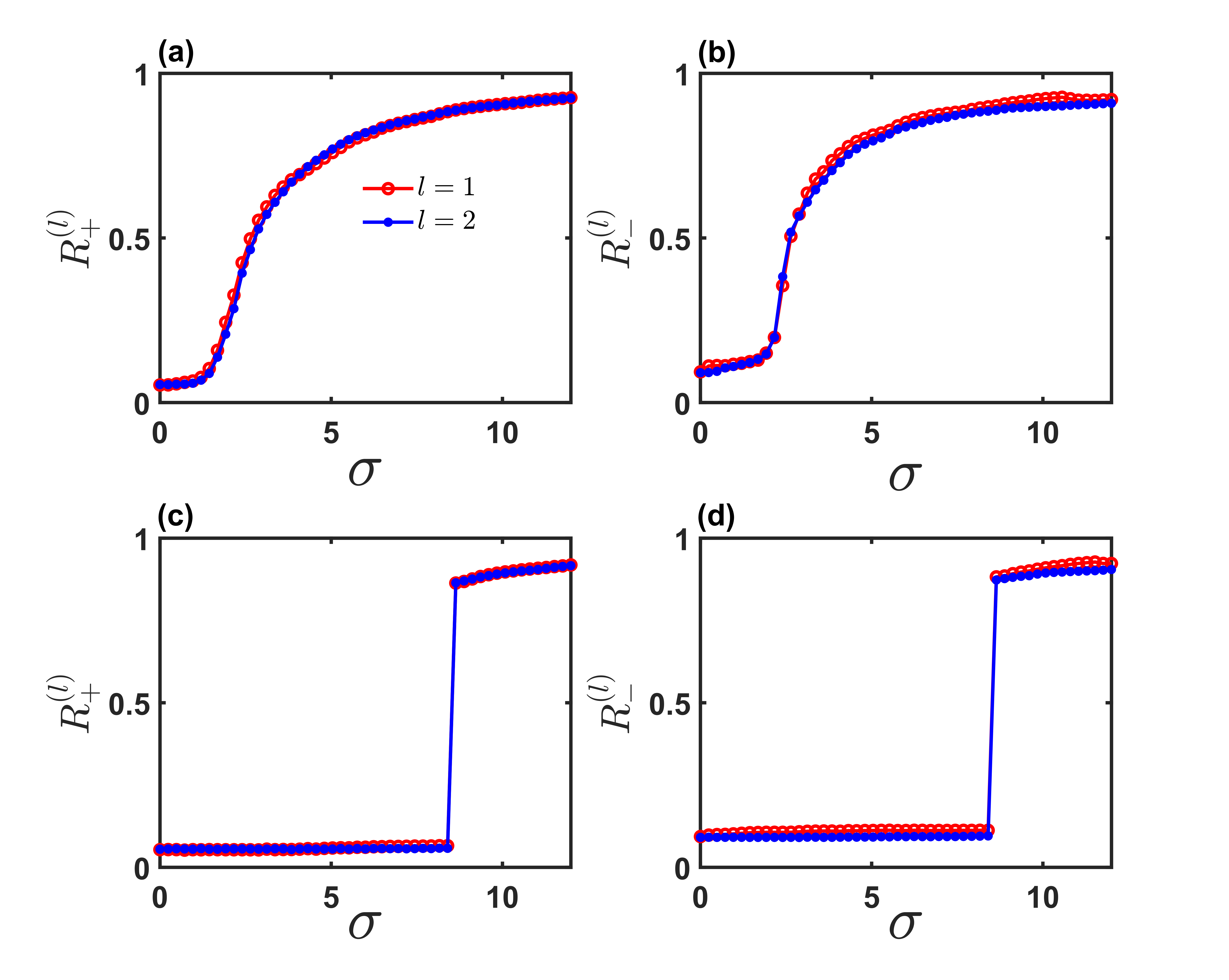}
    \caption{\textbf{Amplitude of the order parameters for projected link phases in the multilayer network with same-dimensional interlayer coupling.} Panels (a) and (c) show the order parameters obtained by projecting link phases onto the triangles of layers $l=1$ and $l=2$, corresponding to models A1 and B1, respectively. Panels (b) and (d) display the order parameters obtained by projecting link phases onto the nodes for models A1 and B2, respectively.}
    \label{fig_b}
\end{figure}
\begin{figure}
    \centering
    \includegraphics[width=1.1\linewidth]{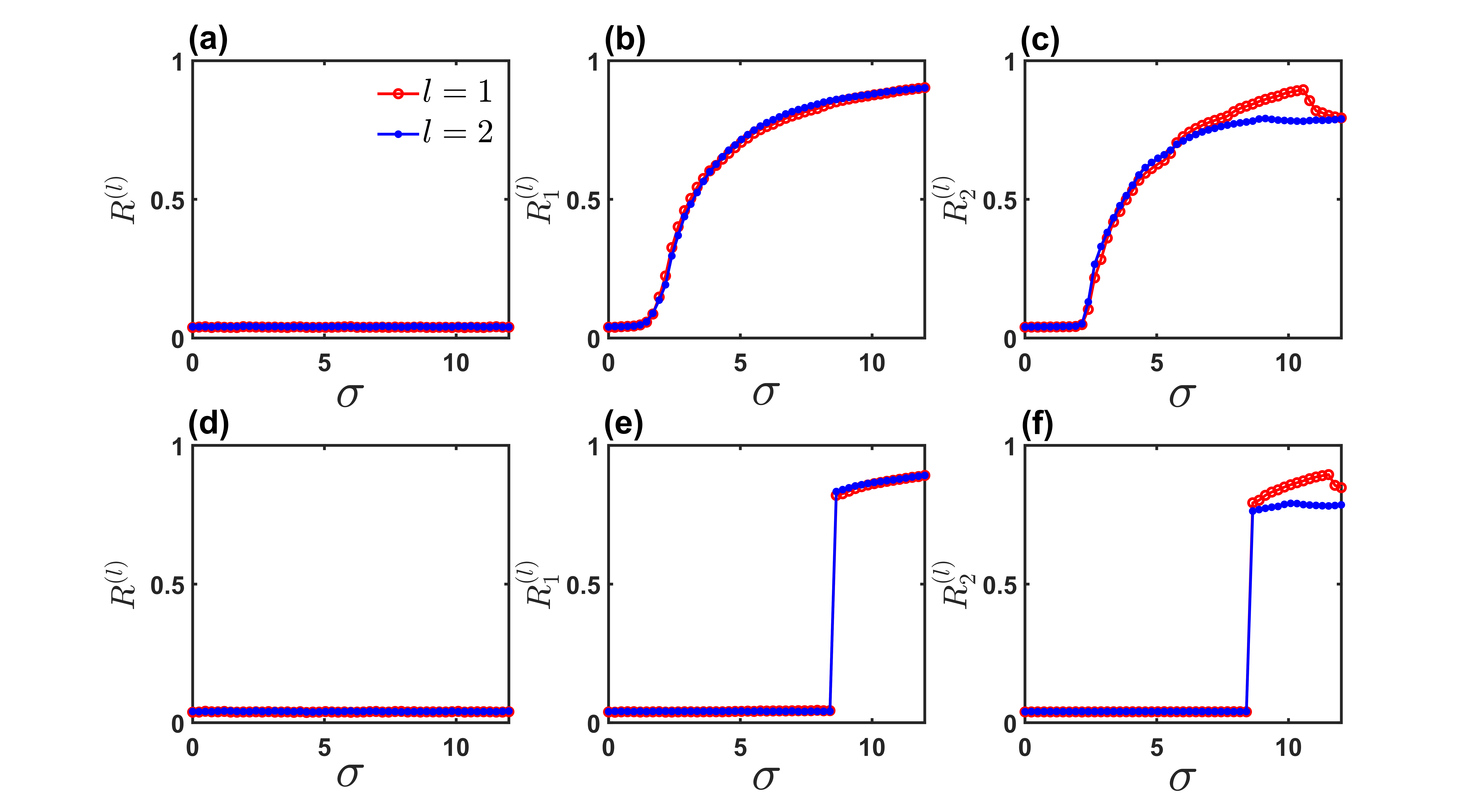}
    \caption{\textbf{Amplitude of the order parameters of the link phases and of their Hodge-decomposed components.} Panels (a,d) display the order parameters of the full link phases, while panels (b,e) and (c,f) show those of the phases sensitive to the solenoidal and irrotational components obtained via the up- and down-Hodge Laplacians, respectively. Results for the uncoupled projections in each layer (model A1) are shown in the upper row, and results for the coupled case (model B1) are shown in the lower row.}
    \label{fig_c}
\end{figure}
\par We simulate \cite{Pal2026Code} these multilayer models given by Eq. \eqref{eq_sd_node} for nodes and Eqs. \eqref{eq_sd_link1} and \eqref{eq_sd_link2} for links considering two configuration models \cite{courtney2016generalized} of simplicial complexes representing the higher-order network structures of the two layers. We set the configuration models of layer-$1$ and layer-$2$ to have $100$ nodes, $516$ links and $278$ triangles, and $100$ nodes, $481$ links and $252$ triangles, respectively. The generalized degree distribution of the nodes in each layer follows a power-law distribution given by \[
P(k) \propto k^{-\gamma},\] where $k$ is the $2$nd degree of the node, i.e., the number of triangles incident to the node, with exponent $\gamma=2.8$. The frequencies of the nodes and links are drawn from two independent standard Cauchy distributions. Figure \ref{fig_a} represents the phase transitions of the  two layers nodes for these multilayer models. Find that both are explosive in nature and there are no significant differences between these two coupling scenarios. The phase transitions of the projected link signal on the nodes and triangles within the two layers are given in Fig. \ref{fig_b}. The transitions of these two link projections are continuous (see Fig. \ref{fig_b} (a, b)) when the link dynamics follows Eq. \eqref{eq_sd_link1}, but are explosive (see Fig. \ref{fig_b} (c, d)) when the link dynamics is Eq. \eqref{eq_sd_link2}. However, the links are not synchronized in this multilayer model (see Fig. \ref{fig_c} (a, d)). The transitions of the amplitudes of the order parameters sensitive to the solenoidal component and irrotational components are both continuous (see Fig. \ref{fig_c} (b, c)) when Eq. \eqref{eq_sd_link1} is the link dynamics, whereas both are explosive (see Fig. \ref{fig_c} (e, f)) when Eq. \eqref{eq_sd_link2} is the dynamics of the link signal. The link-signal dynamics can be decomposed into solenoidal and irrotational components through the Hodge decomposition. These components play distinct and complementary roles in the synchronization process: the solenoidal component drives the synchronization of the link phases projected onto triangle simplices, whereas the irrotational component governs the synchronization of the projected phases onto nodes. Figure \ref{fig_c} shows that the transition to synchronization of the full link dynamics is accompanied by the coherent ordering of both components, thereby clarifying the topological mechanism underlying the observed behavior. Although the two layers are governed by different dynamical equations, we find that this does not result in different types of synchronization transitions: the qualitative behavior observed in Figs. (\ref{fig_a},\ref{fig_b},\ref{fig_c}) is the same for both layers and is robust under exchange of the layer topologies.
\par In this framework of the multilayer network, only the signals of the same-dimensional simplices interact with each other, and the results exhibit the same qualitative behavior as the monolayer network studied by Mill\'an et al. \cite{millan2020explosive}. But these topological signals of different dimensions, e.g., the nodes and triangles, can interact with each other in a network. This interaction can happen either through incorporation of the Dirac operator \cite{bianconi2021topological} or in a trivial way of adaptive coupling \cite{zhang2015explosive}, as we are doing by modulating the coupling strength through the amplitude of the order parameters. Ghorbanchian et al. \cite{ghorbanchian2021higher} used this second method to couple the topological signals of different dimensions in a monolayer network of simplicial complex. In the next section, we also follow the same way to couple these signals in the multilayer network.
\subsection{Multilayer coupling between any dimensional signals} In addition the coupling between the same-dimensional signals, the interaction between the different dimensional signals, for instance, the coupling strength modulation between the node and down-link signals, is also allowed in the multilayer model.
In this framework, we construct a multilayer network in which the nodes evolve as
\begin{equation}
\label{eq_gen_node}
    \dot{\Theta}^{(l)} = \Omega^{(l)} - \sigma f_0^{(l)}(R_0^{(l')},R_-^{l'},R_-^{l})B^{(l)}_1 \sin\left([B^{(l)}_1]^T{\Theta}^{(l)}\right),
\end{equation}
$\textit{where}~l'\neq l\in\{1,2\}.$
As in the previous case, the dynamics of the links in the multilayer network can be characterized by two distinct evolution schemes. Accordingly, we introduce two classes of multilayer frameworks, namely A2 and B2:
\begin{enumerate}
\renewcommand{\labelenumi}{\Alph{enumi}2.}
    \item Nodes-Links-Multilayer (NLM) model. When the projected phases of the links onto the nodes and triangles in the layers are not coupled, the evolution equation is
\begin{equation}
    \begin{split}
\dot{\Phi}^{(l)} = \tilde{\Omega}^{(l)} - \sigma f_1^{(l)}(R_0^{(l')}, R^{(l')}_-,R_0^{(l)}) [B^{(l)}_1]^T \sin\left(\Phi^{(l)}_-\right)\\- \sigma R^{(l')}_+ [B^{(l)}_2] \sin\left(\Phi^{(l)}_+\right). 
\label{eq_gen_link1}
\end{split}
\end{equation}
\item Nodes-Link-Triangles-Multilayer (NLTM) model. When those projected phases are coupled, the differential equation of the evolution of the link signal is
\begin{equation}
    \begin{split}
\dot{\Phi}^{(l)} = \tilde{\Omega}^{(l)} - \sigma f_1^{(l)}(R_0^{(l')},R_0^{(l)}, R^{(l')}_-)R^{(l)}_+ [B^{(l)}_1]^T \sin\left(\Phi^{(l)}_-\right)\\- \sigma R^{(l')}_+R^{(l)}_- [B^{(l)}_2] \sin\left(\Phi^{(l)}_+\right).   
\end{split}
\label{eq_gen_link2}
\end{equation}
\end{enumerate}

\begin{figure}
    \centering
    \includegraphics[width=1.0\linewidth]{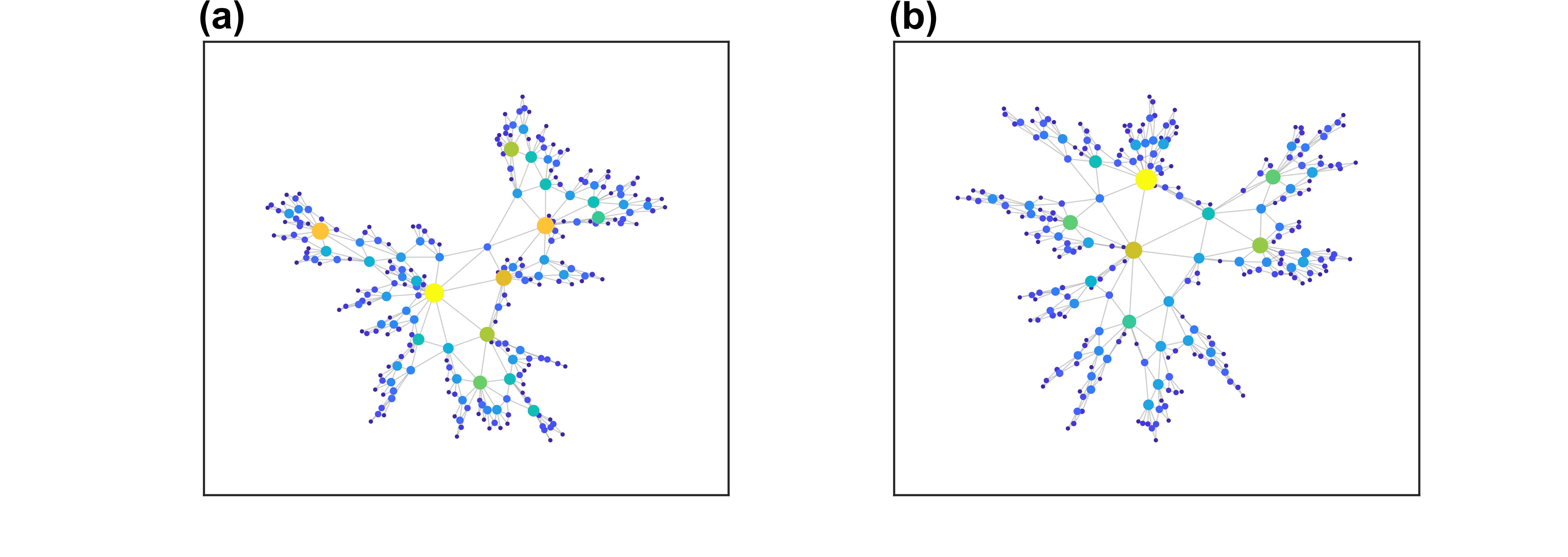}
    \caption{\textbf{The network skeleton of the network geometry with flavor (NGF) model \cite{bianconi2016network} of simplicial complex of dimension $d=2$ and flavor $s=-1$.} (a) and (b) are the network skeletons of $N_0=250$ nodes of layer-$1$, and layer-$2$, respectively.}
    \label{fig_d}
\end{figure}
\begin{figure}
    \centering
    \includegraphics[width=1.1\linewidth]{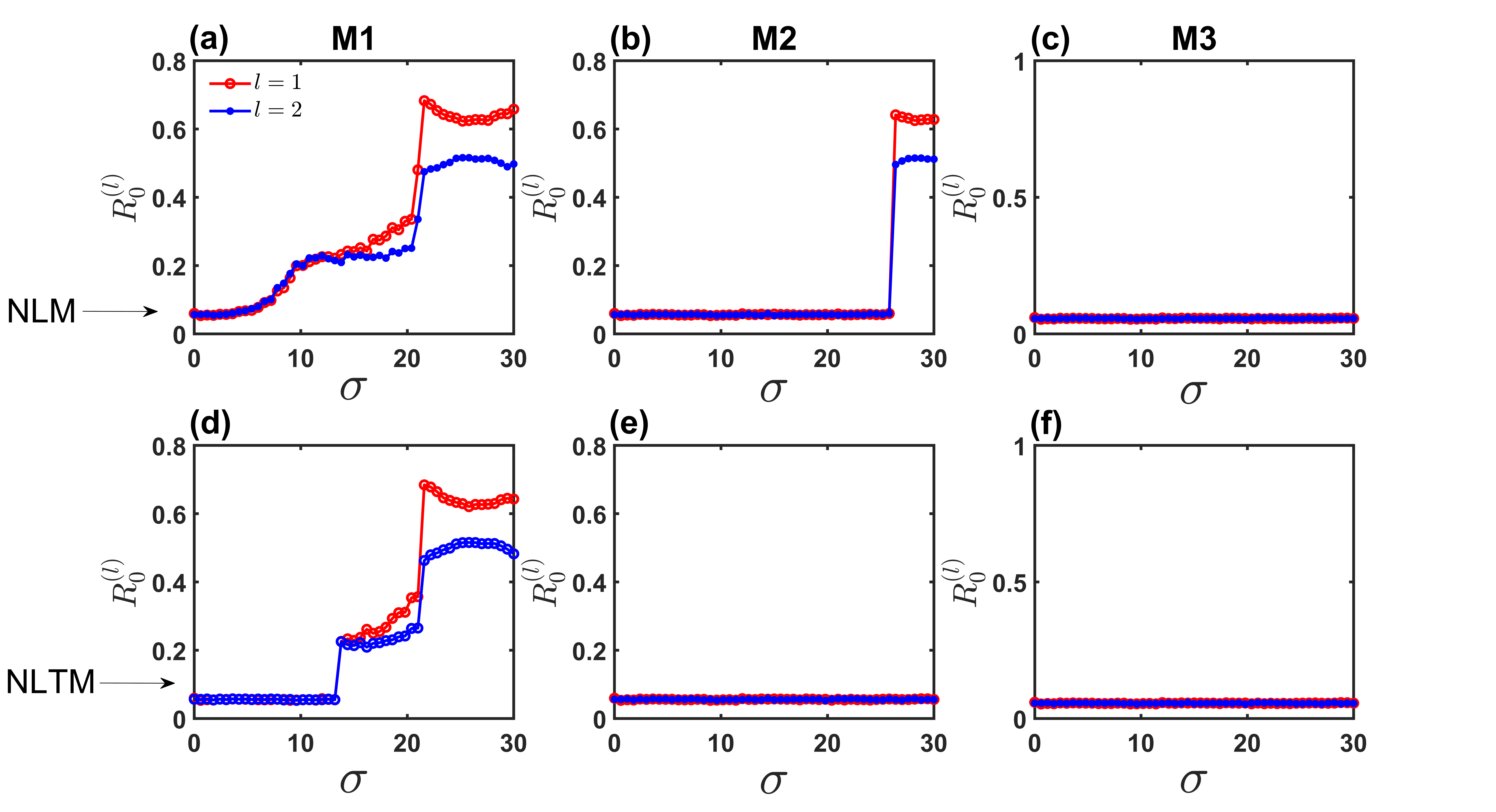}
    \caption{\textbf{Node-signal synchronization transitions in cross-dimensionally coupled models.} Panels (a)–(c) show the order-parameter amplitudes for the multilayer network models M1, M2, and M3 of the NLM class (A2), respectively. Panels (d)–(f) present the corresponding results for the multilayer network models M1, M2, and M3 of the NLTM class (B2). For all cases, the network layers are constructed using the NGF model, and the corresponding network skeletons are shown in Fig.~\ref{fig_d}.}
    \label{fig_e}
\end{figure}
\begin{figure}
    \centering
    \includegraphics[width=1.1\linewidth]{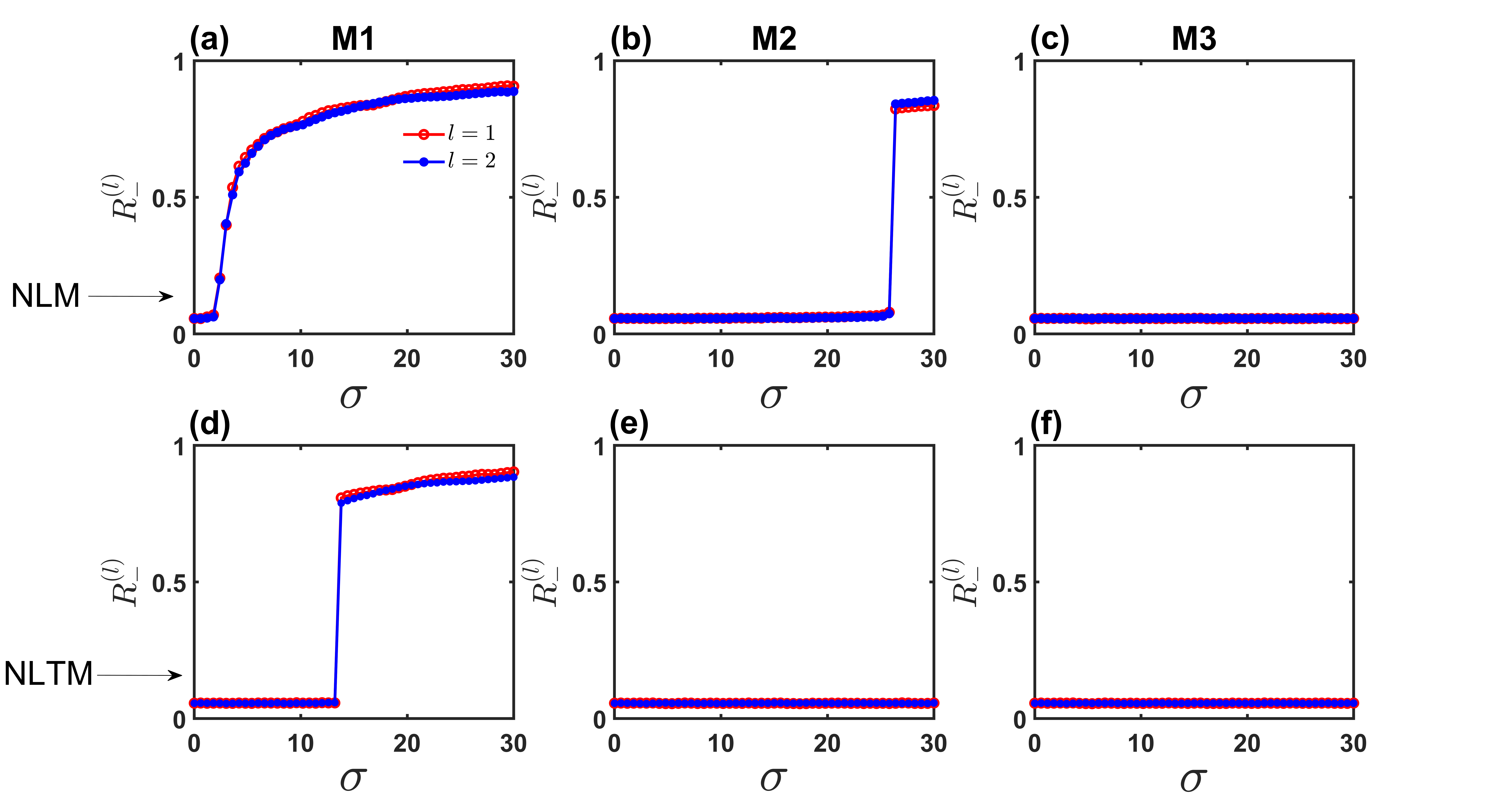}
    \caption{\textbf{Synchronization transition of the down-link signals in the cross-dimensionally coupled models.} Panels (a)–(c) display the synchronization transitions of the down-link signals for the multilayer network models M1, M2, and M3 of the NLM class (A2), respectively. Panels (d)–(f) show the corresponding synchronization transitions for the multilayer network models M1, M2, and M3 of the NLTM class (B2), respectively. The network layers are generated using the NGF model, with skeletons shown in Fig.~\ref{fig_d}.}
    \label{fig_f}
\end{figure}
\begin{figure}
    \centering
    \includegraphics[width=1.1\linewidth]{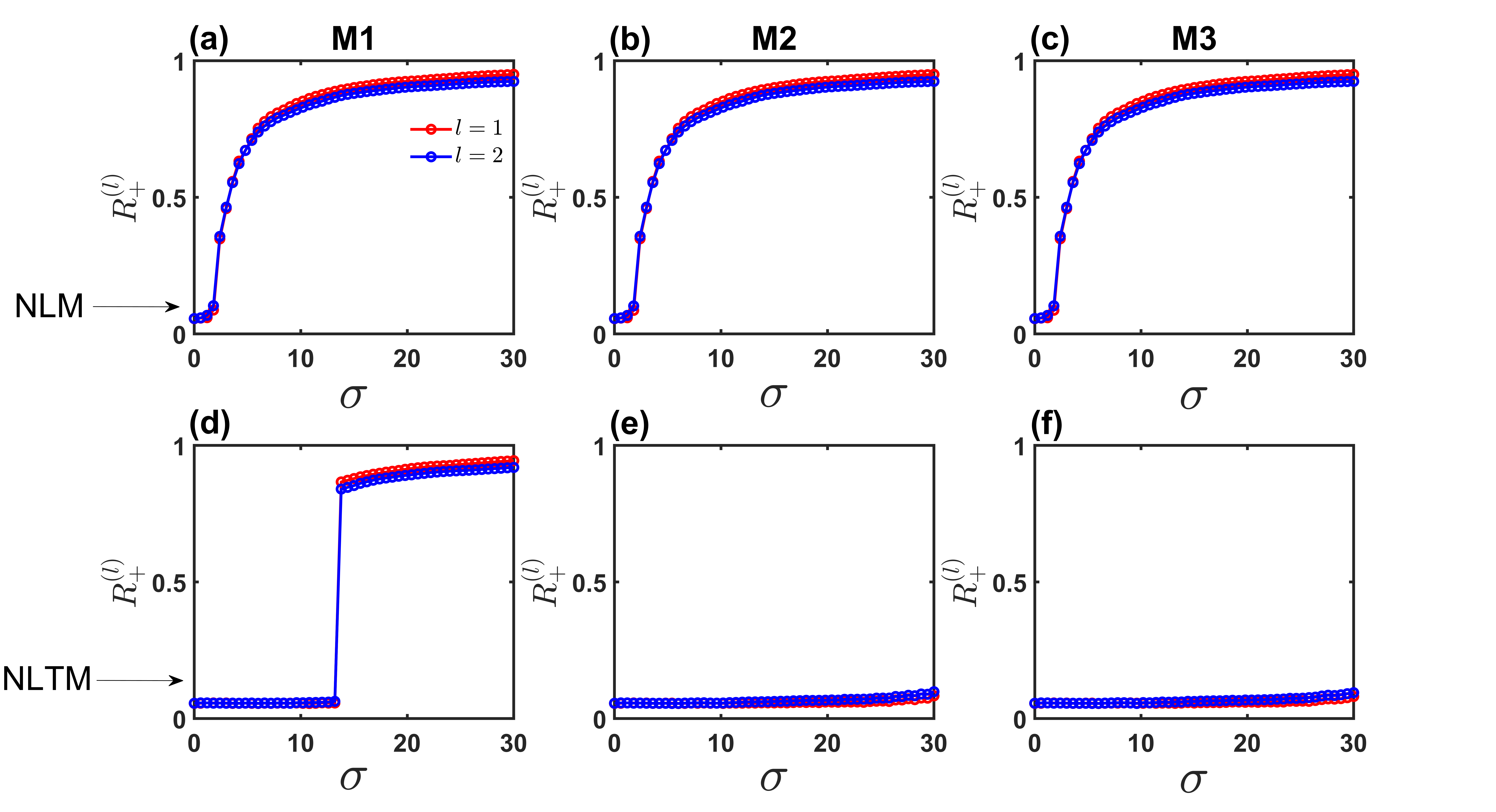}
    \caption{\textbf{Synchronization transitions of link-signal projections onto triangles within each layer of the cross-dimensionally coupled models.} Panels (a)–(c) show the synchronization profiles for the multilayer network models M1, M2, and M3 of the NLM class (A2), respectively. Panels (d)–(f) present the corresponding synchronization profiles for the multilayer network models M1, M2, and M3 of the NLTM class (B2). In all cases, the projections are evaluated on the triangular faces of each layer, whose network topologies are generated using the NGF model; the corresponding network skeletons are shown in Fig.~\ref{fig_d}.}
    \label{fig_g}
\end{figure}
\par
\begin{figure}
     \centering
     \includegraphics[width=1.1\linewidth]{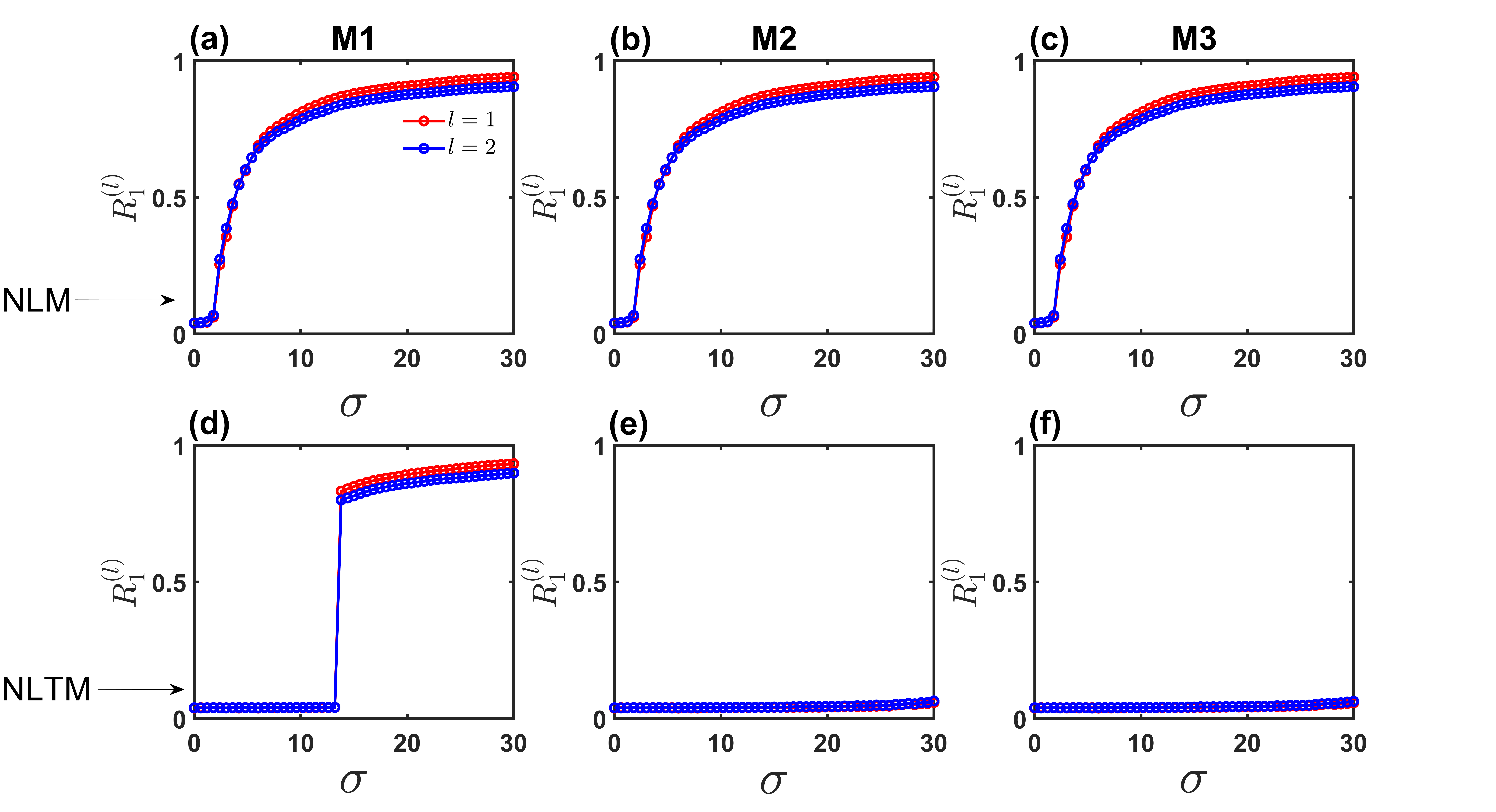}
     \caption{\textbf{Synchronization profile of the phases sensitive to the solenoidal components of the link signals in the cross-dimensionally coupled models.} Panels (a)–(c) display the synchronization transitions for the multilayer network models M1, M2, and M3, respectively, belonging to the NLM class (A2). Panels (d)–(f) show the corresponding synchronization transitions for the multilayer network models M1, M2, and M3 of the NLTM class (B2). The network layers are generated using the NGF model, and their skeletons are shown in Fig.~\ref{fig_d}.}
     \label{fig_h}
 \end{figure}

\begin{figure}
     \centering
     \includegraphics[width=1.1\linewidth]{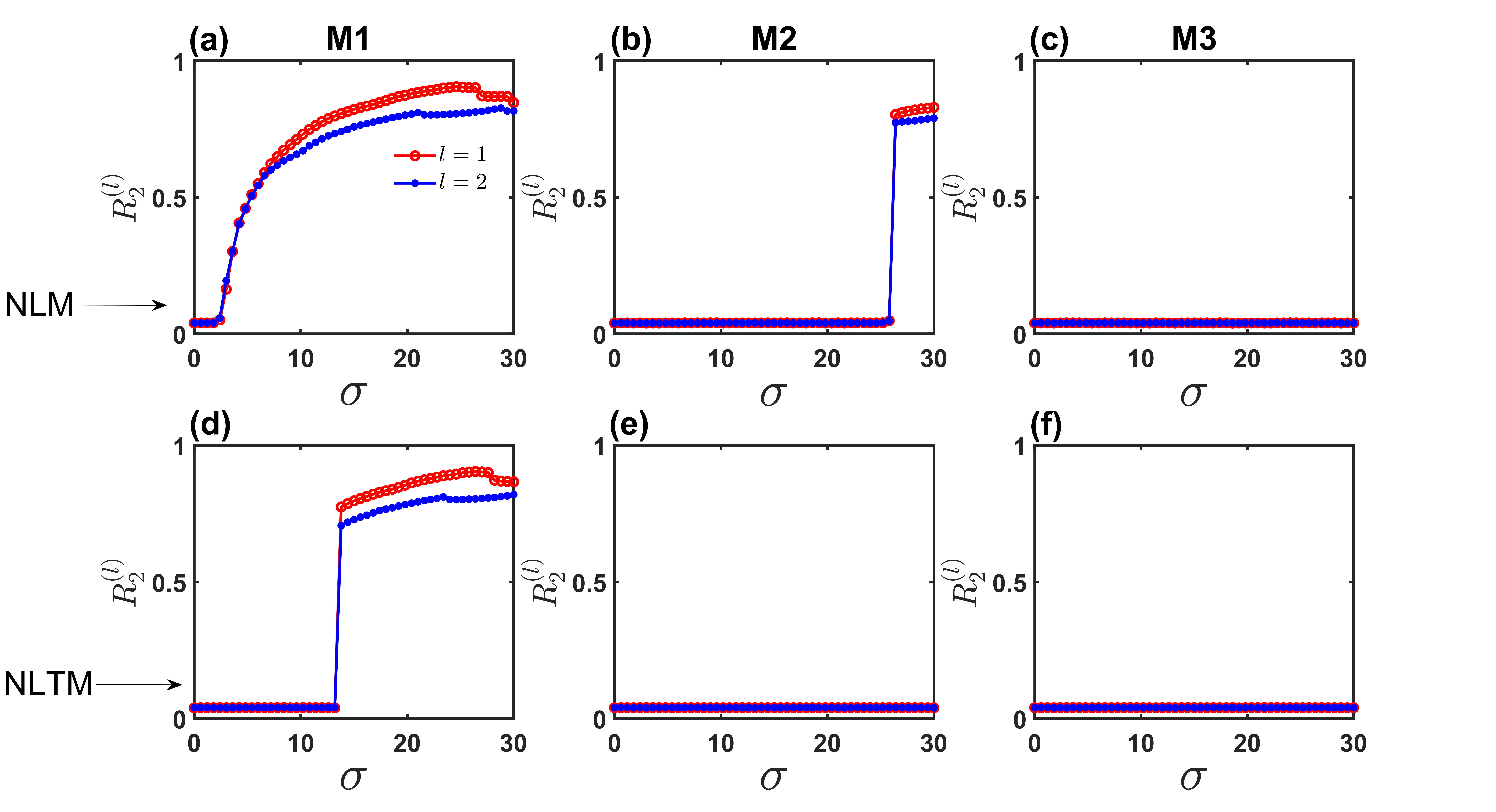}
     \caption{\textbf{Synchronization transition of the phases that are sensitive to the irrotational components of the link signals in the cross-dimensionally coupled models.} The top panels [(a)–(c)] show the synchronization transition profiles for the multilayer networks M1, M2, and M3 belonging to the NLM class (A2), respectively. The bottom panels [(d)–(f)] present the corresponding synchronization transition profiles for the multilayer networks M1, M2, and M3 belonging to the NLTM class (B2), respectively. All network layers are constructed using the NGF model; the corresponding network skeletons are shown in Fig.~\ref{fig_d}.}
     \label{fig_i}
 \end{figure}

 In this multilayer network model, there are two adaptive terms, $f_0^{(l)}$ dependent on three values $R_0^{(l')},R_-^{l'},R_-^{l}$, modulating the coupling strength of the node signals and $f_1^{(l)}$ dependent on $R_0^{(l')},R_0^{(l)}, R^{(l')}_-$, modulating the coupling strength of the link signals in the layer-$l$. In this case, we simulate the dynamical Eqs. (\eqref{eq_gen_node}, \eqref{eq_gen_link1}, \eqref{eq_gen_link2}) of the node and link signals by choosing the NGF models of simplicial complexes for the two layers. We consider two independent realizations of two-dimensional network geometry with flavor (NGF) with flavor $s=-1$, each consisting of $250$ nodes, $497$ links and $248$ triangles. In the NGF model with dimension $d=2$ and flavor $s=-1$, the simplicial complex grows by sequentially adding triangles. At each step, a new triangle is attached to an existing link ($1$-simplex) selected according to a degree-dependent attachment rule. The choice $s=-1$ enforces that each link can belong to at most two triangles, yielding a manifold-like simplicial structure without branching at links. This procedure generates a two-dimensional simplicial complex composed of nodes, links, and triangles with controlled higher-order topology. The network skeletons of the two simplicial complexes corresponding to the two layers are shown in Fig.~\ref{fig_d}. By choosing different forms of these adaptive variables $f_0^{(l)},f_1^{(l)}$ in terms of the amplitudes of the order parameters, we can create several multilayer frameworks. In particular, when we take $f_0^{(l)}=R_0^{(l')}R_-^{l'}R_-^{l}$ and $f_1^{(l)}=R_0^{(l')}R_0^{(l)}R^{(l')}_-$ for the models M3 in the classes NLM and NLTM, we observe that the node phases and the down-link phases do not exhibit clear signatures of synchronization within the range of coupling strength $\sigma$ explored, (see the third column of Figs. \ref{fig_e}, \ref{fig_f}) suggesting that the transition is shifted to larger values of $\sigma$. Although the projections of the link signal on the triangles are getting synchronized continuously (see Fig. \ref{fig_g} (c)). We also see that the order parameter $R_1^{(l)}$ sensitive to the solenoidal components of the links are going from $0$ to $1$ continuously (see Fig. \ref{fig_h} (c)), but $R_2$, which depends on the irrotational components of the links, remains approximately zero over the range of coupling strengths $\sigma$ considered. (see Fig. \ref{fig_i} (c,f)).
\par Interestingly, when we do not take into account the values of $R_0^{(l')},R_0^{(l)}$ in the adaptive term $f_1^{(l)}$, we observe synchronization of the node phases (see first column of Fig. \ref{fig_e}) as well as the synchronization of the up-link (see first column of Fig. \ref{fig_g}) and down-link phases (see first column of Fig. \ref{fig_f}) either in a continuous or in a discontinuous way. In this particular framework, we are considering multilayer higher-order Kuramoto dynamics in which the node signals can interact with the link signals, but the link signals can not. Therefore, we have two particular adaptive multilayer frameworks for the two values of the adaptive term $f_1^{(l)}$, viz. $f_1^{(l)}=R_0^{(l')}R_0^{(l)}R^{(l')}_-$ and $f_1^{(l)}=R^{(l')}_-$ which combined suggest us that node signals effecting on the link signal in the multilayer network has an significant effect on the synchronization profile of each type of phase.
By observing this scenario, we consider three different values of the adaptive term $f_1^{(l)}$ to see how the effect of the node signal on the link signal affects the overall dynamics of the node, up-link, and down-link phases.\\

So, fixing the adaptive term $f_0^{(l)}=R_0^{(l')}R_-^{l'}R_-^{l}$, we construct three multilayer network models, namely M1, M2, and M3 corresponding to the three different values of the adaptive term $f_1^{(l)}$ given as:
\begin{enumerate}
    \item $f_1^{(l)}=R^{(l')}_-$ for the multilayer model M1,
    \item $f_1^{(l)}=R_0^{(l')}R^{(l')}_-$ for the multilayer model M2,
    \item $f_1^{(l)}=R_0^{(l')}R_0^{(l)}R^{(l')}_-$ for the multilayer model M3. 
\end{enumerate}
\par We simulate these three multilayer models, M1, M2, and M3 given by Eq. \eqref{eq_gen_node} for nodes and Eqs. \eqref{eq_gen_link1} and \eqref{eq_gen_link2} for links with the corresponding values of the two adaptive terms $f_0^{(l)},f_1^{(l)}$, by considering two network geometries with flavor (NGFs) \cite{bianconi2016network}, whose network skeletons are in Fig. \ref{fig_d}. The frequencies of the nodes and links are drawn from two independent standard Cauchy distributions. The amplitudes of the order parameters of the node, down-link, up-link, and the order parameter sensitive to the solenoidal and irrotational components of the links in each layer are given separately in Figs. \ref{fig_e}, \ref{fig_f}, \ref{fig_g}, \ref{fig_h}, and \ref{fig_i}. In these figures, the first, second and third columns represent the transitions of the multilayer models M1, M2 and M3, respectively. The rows represent NLM (given by A2) and NLTM (given by B2) versions of each of the three models (M1, M2, and M3).
\par The results of the synchronization transitions of the topological signals for all explored multilayer models shown above highlight the distinct roles of the different adaptive components in shaping the synchronization transitions. First, by comparing the multilayer models M1, M2, and M3 in the two classes, NLM and NLTM, we observe that the presence of node-level adaptation progressively suppresses the synchronization of node signals (see the rows in Fig.~\ref{fig_e}) or of the down-link signals (see the rows in Fig.~\ref{fig_f}) or of the up-link signals (see the rows in Fig.~\ref{fig_g}), leading to delayed or weakened transitions. Second, a comparison between the NLM and NLTM classes across the model M2 reveals that the up-link adaptive term further obstructs synchronization of node signals (see the second column in Fig.~\ref{fig_e}) or of the down-link signals (see the second column in Fig.~\ref{fig_f}) or of the up-link signals (see the second column in Fig.~\ref{fig_g}), significantly reducing coherence in both node signals and the projected link dynamics. Finally, by tracking the transitions of the node signals of model M1 in the NLM class (see Fig.~\ref{fig_e}(a) with the models A1 and B1 in the class of the same dimensionally coupled multilayer framework (see Fig.~\ref{fig_a}, we identify the role of the down-link adaptive term, which also acts to hinder synchronization of the node signal. \\Overall, these results demonstrate that the combined action of node, up-link, and down-link adaptive interactions systematically obstructs the onset of synchronization, emphasizing the nontrivial impact of multi-level adaptive mechanisms on collective dynamics. To gain a deeper understanding of these effects, we now develop a theoretical description of the synchronization dynamics in terms of the amplitudes of the relevant order parameters.

\par In the next section, we derive the theoretical expressions of the amplitude of the order parameters of the node signals and the projection of the link signals on the nodes in each layer. We consider in particular the multilayer model M2 from the NLM class, while the derivation procedure is the same for the other models, M1 and M3, from any class.

 \subsection{Backward transition: analytical treatment}
 In this section, we focus on model M2 belonging to the NLM class of multilayer models. The analytical treatment presented in this section follows the strategy developed for the single-layer case in \cite{ghorbanchian2021higher}. 
 Here, we reproduce the main steps of that derivation and extend them to the present multilayer setting, where interlayer adaptive couplings and layer-dependent order parameters introduce additional self-consistency conditions.
 We derive analytical results for the synchronization transition of the corresponding topological signals and compare them with numerical simulations. While the formal structure of the analysis remains the same, the multilayer nature of the system leads to modified effective couplings and coupled self-consistency equations that are absent in the single-layer case. We focus on the backward (desynchronization) transition, which is obtained by decreasing the coupling strength from the synchronized state. Accordingly, the following self-consistency analysis characterizes the backward synchronization transition. All analytical calculations in this section are performed in the thermodynamic limit $N_0 \to \infty$, where sums over nodes can be replaced by ensemble averages over the frequency distribution.
 The two adaptive terms of this multilayer model are  $f_0^{(l)}=R_0^{(l')}R_{-}^{(l)}R_{-}^{(l')},~f_1^{(l)}=R_0^{(l')} R_{-}^{(l')}$ for $l,l' \in \{1,2\}$. \\In the layer $l$, the dynamical equations of the node phases are 
\begin{equation}
\begin{split}
 \dot{\theta}^{(l)}_i = \omega^{(l)}_i + \sigma R_0^{(l')}R_{-}^{(l)}R_{-}^{(l')}\sum_{j=1}^{N_0}a_{ij}^{(l)}\sin\left(\theta^{(l)}_j-\theta^{(l)}_i\right)
 \quad l' \neq l.
\end{split}   
\end{equation}
Here, we rewrite this equation in terms of the full stacked vector $\Theta$, representing the node topological signal (i.e., the $0$-cochain), rather than in its component-wise form given by the individual entries $\theta_i$, which describes the evolution of the node signals, 
\begin{equation}
    \dot{\Theta}^{(l)} = \Omega^{(l)} - \sigma R_0^{(l')}R_{-}^{(l)}R_{-}^{(l')}B_1^{(l)} \sin\left([B_1^{(l)}]^\top\Theta^{(l)}\right) \quad l' \neq l.
\end{equation}

The link signal $\Phi^{(l)}$ has the following dynamics:
\begin{equation}
\begin{split}
    \dot{\Phi}^{(l)} = \tilde{\Omega}^{(l)} &- \sigma R_0^{(l')} R_{-}^{(l')} [B_1^{(l)}]^\top \sin\left(\Phi_{-}^{(l)}\right) \\
    &- \sigma R_{+}^{(l')} [B_2^{(l)}] \sin\left(\Phi_{+}^{(l)}\right).
\end{split}
\end{equation}
Therefore, its projections on the nodes $\Phi^{(l)}_-=B_1^{(l)}\Phi^{(l)}$, in the individual layer, have dynamics
\begin{equation}
\begin{split}
    \dot{\Phi}^{(l)}_- = \hat{\Omega}^{(l)} &- \sigma R_0^{(l')} R_{-}^{(l')}L_0^{(l)} \sin\left(\Phi_{-}^{(l)}\right),
\end{split}
\end{equation} where $\hat{\Omega}^{(l)}=B_1^{(l)}\tilde{\Omega}^{(l)}$.
The dynamics of its $i$-th component $\psi_i^{(l)}$ of the vector $\Phi^{(l)}_-$ follow the equations
\begin{equation}
\begin{split}
    \dot{\psi}^{(l)}_i = \hat{\omega}^{(l)}_i &+ \sigma R_0^{(l')} R_{-}^{(l')} \sum_{j=1}^{N_0}a_{ij}^{(l)}\left(\sin(\psi_{j}^{(l)})-\sin(\psi_{i}^{(l)})\right)\\
    =\hat{\omega}^{(l)}_i &+ \sigma R_0^{(l')} R_{-}^{(l')} \sum_{j=1}^{N_0}a_{ij}^{(l)}\sin(\psi_{j}^{(l)})\\&-\sigma R_0^{(l')}R_{-}^{(l')}k^{(l)}_i\sin(\psi_{i}^{(l)}).
\end{split}
\end{equation}

\par We take an approximation of the 1-skeletons in each layer in the theoretical analysis. So, we replace adjacency matrices with their expectations in uncorrelated network ensembles as,
\begin{equation}
    a_{ij}^{(l)} \rightarrow \frac{k_i^{(l)}k_j^{(l)}}{\langle K^{(l)} \rangle N_0},
\end{equation} where $k_i^{(l)}$ is the degree of $i$-node in layer-$l$, and $\langle K^{(l)} \rangle$ is their average degree.
Before proceeding, we state the assumptions underlying the annealed approximation used in this section. 
We assume that the network skeletons of the simplicial complexes in each layer satisfy the structural cutoff condition
\begin{equation}
k_i^{(l)} \ll \sqrt{\langle K^{(l)} \rangle N_0}, \qquad \forall i,l,
\end{equation}
which suppresses degree--degree correlations and allows the replacement of the discrete network structure by its ensemble average.
Under this condition, the annealed approximation is expected to be valid for both sparse and dense networks.

\par Following \cite{ghorbanchian2021higher}, we introduce the complex order parameters,
\begin{align}
    \hat{R}_0^{(l)} e^{\text{i}\hat{\Theta}^{(l)}} &= \sum_{j=1}^{N_0} \frac{k^{(l)}_j}{\langle K^{(l)}\rangle N_0} e^{\text{i}\theta_j^{(l)}}, \\
    \hat{R}_{-}^{(l)} e^{\text{i}\hat{\Psi}^{(l)}} &= \sum_{j=1}^{N_0} \frac{k^{(l)}_j}{\langle K^{(l)}\rangle N_0} e^{\text{i}\psi_j^{(l)}}.
\end{align}
The node dynamics in the annealed approximation is therefore,

\begin{equation}
 \dot{\Theta}^{(l)} = \Omega^{(l)} - \sigma R_0^{(l')}R_{-}^{(l)}R_{-}^{(l')}\hat{R}_0^{(l)}K^{(l)}\odot\sin\left(\Theta^{(l)}-\hat{\Theta}^{(l)}\right)
 \end{equation}
 i.e.,
 \begin{equation}
  \dot{\theta}^{(l)}_i = \omega^{(l)}_i - \sigma R_0^{(l')}R_{-}^{(l)}R_{-}^{(l')}\hat{R}_0^{(l)}k^{(l)}_i\sin\left(\theta^{(l)}_i-\hat{\Theta}^{(l)}\right),
\end{equation}
for $i=1,2,\dots,N_0.$ \\Substituting $\theta^{(l)}_i-\hat{\Theta}^{(l)}=\delta\theta_i^{(l)}$ and $\dot{\hat{\Theta}}^{(l)}=\Omega_0^{(l)}$, the evolution equation of the nodes becomes
\begin{equation}
\label{pertur}
  \delta\dot{\theta}^{(l)}_i = \omega^{(l)}_i-\Omega_0^{(l)} - \sigma R_0^{(l')}R_{-}^{(l)}R_{-}^{(l')}\hat{R}_0^{(l)}k^{(l)}_i\sin\left(\delta\theta^{(l)}_i\right). 
\end{equation} 
The fixed point of this perturbation Eq. \eqref{pertur} is defined by $\sin{(\delta\theta^{(l)}_i)}=\frac{\omega^{(l)}_i-\Omega_0^{(l)}}{\sigma R_0^{(l')}R_{-}^{(l)}R_{-}^{(l')}\hat{R}_0^{(l)}k^{(l)}_i} \equiv c_i^{(l)}$, where the oscillators satisfying the inequality\\ $|\omega^{(l)}_i-\Omega_0^{(l)}|\le \sigma R_0^{(l')}R_{-}^{(l)}R_{-}^{(l')}\hat{R}_0^{(l)}k^{(l)}_i$ defined to be phase-locked. The order parameters become
\begin{equation}
    \hat{R}_0^{(l)} = \sum_{|c_j^{(l)}|\le 1} \frac{k^{(l)}_j}{\langle K^{(l)}\rangle N_0} \cos{(\delta \theta_j^{(l)})},
 \end{equation}
 since the drifting oscillators do not contribute to this sum \cite{zhang2015explosive,ji2013cluster}. So the order parameters of the node signal with and without considering the annealed approximation are as follows,
 \begin{widetext}
       \begin{equation}
    \begin{split}
    \hat{R}_0^{(l)} &= \sum_{j=1}^{N_0} \frac{k^{(l)}_j}{\langle K^{(l)}\rangle N_0}\int_{|c_j^{(l)}|\le 1}g^{(l)}(\omega_j^{(l)})d \omega_j^{(l)}\sqrt{1 - \left(\frac{\omega_j^{(l)} - \Omega_0^{(l)}}{\sigma R_0^{(l')}R_{-}^{(l)}R_{-}^{(l')}k_j^{(l)}\hat{R}_0^{(l)}}\right)^2}\\
   R_0^{(l)} &= \frac{1}{ N_0} \sum_{j=1}^{N_0} \int_{|c_j^{(l)}|\le 1}g^{(l)}(\omega_j^{(l)})d \omega_j^{(l)}\sqrt{1 - \left(\frac{\omega_j^{(l)} - \Omega_0^{(l)}}{\sigma R_0^{(l')}R_{-}^{(l)}R_{-}^{(l')}k_j^{(l)}\hat{R}_0^{(l)}}\right)^2},
    \end{split}
    \label{node_op}
    \end{equation}  
 \end{widetext}
where $g^{(l)}$ is the distribution function of the natural frequencies of the node phases in the $l$-th layer.

Now, in the annealed approximation, the down-link dynamics, that is, the dynamics of the projections of the link signal on the nodes in the individual layers, are
\begin{widetext}
\begin{equation}
    \begin{split}
        \dot{\psi}_i^{(l)} =\hat{\omega}^{(l)}_i + \sigma R_0^{(l')} R_{-}^{(l')}k_i^{(l)} \hat{R}_-^{(l)}\sin(\hat{\Psi}^{(l)})-\sigma R_0^{(l')} R_{-}^{(l')}k^{(l)}_i\sin(\psi_{i}^{(l)}).
    \end{split}
    \label{node_op2}
\end{equation}
\end{widetext}
The above Eq. \eqref{node_op2} can be written in vector form as,
\begin{widetext}
\begin{equation}
    \begin{split}
        \dot{\Phi}_-^{(l)} =\hat{\Omega}^{(l)} + \sigma R_0^{(l')} R_{-}^{(l')} \hat{R}_-^{(l)}\sin(\hat{\Psi}^{(l)})K^{(l)}-\sigma R_0^{(l')} R_{-}^{(l')}K^{(l)}\odot\sin(\Phi_{-}^{(l)}).
    \end{split}
\end{equation}
\end{widetext}
Using the Ott-Antonsen ansatz \cite{ott2008low} for the probability density, we are going to derive the expression of $\hat{R}_-^{(l)}$ and $ R^{(l)}_-$. The probability density function $\rho^{(l)}_j(\psi^{(l)},t|\hat{\omega}^{(l)})$ that the node $j$ of the layer $l$ is associated with a projected phase of the link equal to $\psi^{(l)}$,
\begin{equation}
\begin{split}
       \rho^{(l)}_j(\psi^{(l)},t|\hat{\omega}^{(l)})=\frac{1}{2\pi}\big[1 + \sum_{m=1}^\infty \big\{(b_j^{(l)}(\hat{\omega}_j^{(l)},t))^m e^{\text{i} m\psi^{(l)}} \\ +\text{c.c.}\big\}\big], 
\end{split}
\end{equation} 
obeys the continuity equation,
\begin{equation}
     \label{continuity}\partial_t\rho^{(l)}_j(\psi^{(l)},t|\hat{\omega}^{(l)})+\partial_{\psi^{(l)}}[\rho^{(l)}_j(\psi^{(l)},t|\hat{\omega}^{(l)})v_j^{(l)}]=0,
\end{equation} with the associated velocity $v_j^{(l)}$ given by $v_j^{(l)}=\kappa^{(l)}_j-\sigma R_0^{(l')} R_{-}^{(l')}k^{(l)}_j\sin(\psi^{(l)})$, where $\kappa_j^{(l)} = \hat{\omega}_j^{(l)} + \sigma R_0^{(l')}R_{-}^{(l')}k_j^{(l)}\hat{R}_{-}^{(l)}\sin\hat{\Psi}^{(l)}$. \\Substituting this expression $v_j^{(l)}$ into the above continuity equation (Eq. \eqref{continuity}) and equating the coefficient of $e^{\text{i}\psi^{l}}$ to zero, we obtain the equation for $b_j^{(l)}$,
\begin{equation}
    \partial_t b_j^{(l)} + \text{i} b_j^{(l)}\kappa_j^{(l)} + \frac{\sigma R_0^{(l')}R_{-}^{(l')}k_j^{(l)}}{2}\left([b_j^{(l)}]^2 - 1\right) = 0.
\end{equation}

For stationary solutions $\partial_t b_j^{(l)} = 0$, it implies
\begin{equation}
    b_j^{(l)} = -\text{i}d_j^{(l)} \pm \sqrt{1 - [d_j^{(l)}]^2},
\end{equation}
where
\begin{equation}
    d_j^{(l)} = \frac{\hat{\omega}_j^{(l)}}{\sigma R_0^{(l')}R_{-}^{(l')}k_j^{(l)}} + \hat{R}_{-}^{(l)}\sin(\hat{\Psi}^{(l)}).
\end{equation}

Finally, the order parameter equations associated with the link signals become,
\begin{align}
    \hat{R}_{-}^{(l)}\cos(\hat{\Psi}^{(l)}) &= \sum_{i=1}^{N_0}\frac{k_i^{(l)}}{\langle K^{(l)} \rangle N_0} \sqrt{1 - [d_i^{(l)}]^2} ~\mathbf{H}(1 - [d_i^{(l)}]^2), \\
    \hat{R}_{-}^{(l)}\sin(\hat{\Psi}^{(l)}) &= \sum_{i=1}^{N_0} \frac{k_i^{(l)}}{\langle K^{(l)} \rangle N_0} \left(\sqrt{[d_i^{(l)}]^2 - 1}~\chi(d_i^{(l)}) + d_i^{(l)}\right),\\
    R_{-}^{(l)}\cos(\Psi^{(l)}) &= \frac{1}{N_0}\sum_{i=1}^{N_0} \sqrt{1 - [d_i^{(l)}]^2} ~\mathbf{H}(1 - [d_i^{(l)}]^2), \\
    R_{-}^{(l)}\sin(\Psi^{(l)}) &= \frac{1}{N_0}\sum_{i=1}^{N_0}  \left(\sqrt{[d_i^{(l)}]^2 - 1}~\chi(d_i^{(l)}) + d_i^{(l)}\right),
\end{align}
where $\chi(d_i^{(l)}) = -\mathbf{H}(d_i^{(l)}-1) + \mathbf{H}(-d_i^{(l)}-1)$, $\mathbf{H}$ is indicating a Heaviside step function. 
The order parameters of the projected link dynamics in the annealed approximation are given by,
\begin{subequations}
  \label{link_op} 
  \begin{widetext}
\begin{align}
  \begin{split}
     \hat{R}_{-}^{(l)} \cos \hat{\Psi}^{(l)} &= \sum_{j=1}^{N_0} \frac{k_j^{(l)}}{\langle K^{(l)} \rangle N_0} \int_{|d_j^{(l)}| \leq 1} d\hat{\omega}_j^{(l)} \, G_j^{(l)}(\langle \hat{\omega}_j^{(l)}\rangle+\hat{\omega}_j^{(l)}) \sqrt{1 - \left( \frac{\hat{\omega}_j^{(l)}}{\sigma R_0^{(l')}R_{-}^{(l')}k_j^{(l)}} + \hat{R}_{-}^{(l)}\sin(\hat{\Psi}^{(l)}) \right)^2},  \\
    \hat{R}_{-}^{(l)} \sin \hat{\Psi}^{(l)} &= -\sum_{j=1}^{N_0} \frac{k_j^{(l)}}{\langle K^{(l)} \rangle N_0} \int_{d_j^{(l)} > 1} d\hat{\omega}_j^{(l)} \, G_j^{(l)}(\langle \hat{\omega}_j^{(l)}\rangle+\hat{\omega}_j^{(l)}) \sqrt{\left( \frac{\hat{\omega}_j^{(l)}}{\sigma R_0^{(l')}R_{-}^{(l')}k_j^{(l)}} + \hat{R}_{-}^{(l)}\sin(\hat{\Psi}^{(l)}) \right)^2 - 1}  \\
    &\quad + \sum_{j=1}^{N_0} \frac{k_j^{(l)}}{\langle K^{(l)} \rangle N_0} \int_{d_j^{(l)} < -1} d\hat{\omega}_j^{(l)} \, G_j^{(l)}(\langle \hat{\omega}_j^{(l)}\rangle+\hat{\omega}_j^{(l)}) \sqrt{\left( \frac{\hat{\omega}_j^{(l)}}{\sigma R_0^{(l')}R_{-}^{(l')}k_j^{(l)}} + \hat{R}_{-}^{(l)}\sin(\hat{\Psi}^{(l)}) \right)^2 - 1}  \\
    &\quad + \sum_{j=1}^{N_0} \frac{k_j^{(l)}}{\langle K^{(l)} \rangle N_0} \int_{-\infty}^{\infty} d\hat{\omega}_j^{(l)} \, G_j^{(l)}(\langle \hat{\omega}_j^{(l)}\rangle+\hat{\omega}_j^{(l)}) \left( \frac{\hat{\omega}_j^{(l)}}{\sigma R_0^{(l')}R_{-}^{(l')}k_j^{(l)}} + \hat{R}_{-}^{(l)}\sin(\hat{\Psi}^{(l)}) \right), 
    \end{split}  
\end{align}
\end{widetext}
\newpage
\begin{widetext}
 \begin{align}
    \begin{split}
        R_{-}^{(l)} \cos \Psi^{(l)} &= \sum_{j=1}^{N_0} \frac{1}{N_0} \int_{|d_j^{(l)}| \leq 1} d\hat{\omega}_j^{(l)} \, G_j^{(l)}(\langle \hat{\omega}_j^{(l)}\rangle+\hat{\omega}_j^{(l)}) \sqrt{1 - \left( \frac{\hat{\omega}_j^{(l)}}{\sigma R_0^{(l')}R_{-}^{(l')}k_j^{(l)}} + \hat{R}_{-}^{(l)}\sin(\hat{\Psi}^{(l)}) \right)^2}, \\
    R_{-}^{(l)} \sin \Psi^{(l)} &= -\sum_{j=1}^{N_0} \frac{1}{N_0} \int_{d_j^{(l)} > 1} d\hat{\omega}_j^{(l)} \, G_j^{(l)}(\langle \hat{\omega}_j^{(l)}\rangle+\hat{\omega}_j^{(l)}) \sqrt{\left( \frac{\hat{\omega}_j^{(l)}}{\sigma R_0^{(l')}R_{-}^{(l')}k_j^{(l)}} + \hat{R}_{-}^{(l)}\sin(\hat{\Psi}^{(l)}) \right)^2 - 1} \\
    &\quad + \sum_{j=1}^{N_0} \frac{1}{N_0} \int_{d_j^{(l)} < -1} d\hat{\omega}_j^{(l)} \, G_j^{(l)}(\langle \hat{\omega}_j^{(l)}\rangle+\hat{\omega}_j^{(l)}) \sqrt{\left( \frac{\hat{\omega}_j^{(l)}}{\sigma R_0^{(l')}R_{-}^{(l')}k_j^{(l)}} + \hat{R}_{-}^{(l)}\sin(\hat{\Psi}^{(l)}) \right)^2 - 1}  \\
    &\quad + \sum_{j=1}^{N_0} \frac{1}{N_0} \int_{-\infty}^{\infty} d\hat{\omega}_j^{(l)} \, G_j^{(l)}(\langle \hat{\omega}_j^{(l)}\rangle+\hat{\omega}_j^{(l)}) \left( \frac{\hat{\omega}_j^{(l)}}{\sigma R_0^{(l')}R_{-}^{(l')}k_j^{(l)}} + \hat{R}_{-}^{(l)}\sin(\hat{\Psi}^{(l)}) \right), 
    \end{split}
\end{align}   
\end{widetext}
\end{subequations}

where $G_j^{(l)}$ is the frequency distribution function of the down-link signal in $l$-th layer.
\subsection{Steady-state solution of the annealed equations}
We assume that the distribution functions $g^{(l)}(\omega)$, and $G^{(l)}_j(\hat{\omega})$ are unimodal and symmetric about their means. Then, setting $\Psi^{(l)}=\hat{\Psi}^{(l)}=0$, and considering the change of variables $\frac{\omega_j^{(l)} - \Omega_0^{(l)}}{\sigma R_0^{(l')}R_{-}^{(l)}R_{-}^{(l')}k_i^{(l)}\hat{R}_0^{(l)}}=x$, we can write Eq. \eqref{node_op} as 
\begin{widetext}
\begin{equation}
\label{node_ann}
    \begin{split}
        \frac{\sigma R_0^{(l')}R_-^{(l')}R_-^{(l)}}{\langle K^{(l)}\rangle N_0}\sum_{j=1}^{N_0}(k_j^{(l)})^2\int_{-1}^{1}g^{(l)}(\Omega_0^{(l)}+x\sigma k_j^{(l)}R_0^{(l')}R_-^{(l')}R_-^{(l)}\hat{R}_0^{l})\sqrt{1-x^2}dx=1,\\
    R_0^{(l)}=\frac{\sigma R_0^{(l')}R_-^{(l')}R_-^{(l)}\hat{R}_0^{l}}{N_0}\sum_{j=1}^{N_0}k_j^{(l)}\int_{-1}^{1}g^{(l)}(\Omega_0^{(l)}+x\sigma k_j^{(l)}R_0^{(l')}R_-^{(l')}R_-^{(l)}\hat{R}_0^{l})\sqrt{1-x^2}dx  
    \end{split}
\end{equation}
\end{widetext}
Using the same approach, by considering the change of variable $y=\frac{\hat{\omega}_j^{(l)}}{\sigma R_0^{(l')}R_{-}^{(l')}k_j^{(l)}}$, we can write Eq. \eqref{link_op} as 
\begin{widetext}
  \begin{equation}
\label{link_down_ann}
    \begin{split}
             \hat{R}_{-}^{(l)} &= \sigma R_0^{(l')}R_{-}^{(l')}\sum_{j=1}^{N_0} \frac{(k_j^{(l)})^2}{\langle K^{(l)} \rangle N_0} \int_{-1}^{1} dy \, G_j^{(l)}(\langle \hat{\omega}_j^{(l)}\rangle+y\sigma R_0^{(l')}R_{-}^{(l')}k_j^{(l)}) \sqrt{1 - y^2},\\
             R_{-}^{(l)} &= \sigma R_0^{(l')}R_{-}^{(l')}\sum_{j=1}^{N_0} \frac{k_j^{(l)}}{N_0} \int_{-1}^{1} dy \, G_j^{(l)}(\langle \hat{\omega}_j^{(l)}\rangle+y\sigma R_0^{(l')}R_{-}^{(l')}k_j^{(l)}) \sqrt{1 - y^2}.
    \end{split}
\end{equation}  
\end{widetext}
The steady-state behavior of the system is determined by the self-consistent relations given in Eqs.~\eqref{node_ann} and \eqref{link_down_ann}, which define the order parameters $R_0^{(l)}$ and $R_-^{(l)}$ implicitly. These equations are interpreted as implicit self-consistency conditions in the annealed approximation. Next, we will deal with the theoretical derivations for the globally coupled multilayer network.
For a fully connected network with $N_0$ nodes per layer, we make the following approximations:
\begin{itemize}
\item Degrees: $k_j^{(l)} = N_0-1 \approx N_0$ (for large $N_0$).
\item Average degree: $\langle K^{(l)}\rangle = N_0-1 \approx N_0$.
\item Coupling rescaling: $\sigma \rightarrow \sigma/(N_0-1)$.
\end{itemize}
We observe that the equations \eqref{node_ann} and \eqref{link_down_ann} for $R_0^{(l)}$, $\hat{R}_0^{(l)}$, and $R^{(l)}_-$ are independent of the
order parameter $\hat{R}^{(l)}_-$. This allows us to derive a fully analytical solution to the model
without solving the equation for $\hat{R}^{(l)}_-$. These equations are influenced by the
distribution $g^{(l)}(\omega^{(l)})$ as well as the set of marginal distributions $G^{(l)}_i(\hat{\omega}_i^{(l)})$.
Nevertheless, it can be demonstrated that as long as the quantities $\frac{\langle (K^{(l)})^2 \rangle}{\langle K^{(l)}\rangle}$ remain finite, the system does not approach the trivial solution 
$R^{(l)}_0 = \hat{R}_0^{(l)} = R^{(l)}_- = 0$ for any finite values of $\sigma$.
We will now show that the nontrivial (unstable) branch of the solution tends toward the trivial solution
only in the limit $\sigma \to \infty$.

\subsection{Marginal distribution in globally coupled network}
As discussed in the previous section, the steady-state behavior of the system is governed by the implicit self-consistency relations in Eqs.~\eqref{node_ann} and \eqref{link_down_ann}, which depend not only on the natural frequency distribution $g^{(l)}(\omega^{(l)})$, but also on the marginal distributions $G_i^{(l)}(\hat{\omega}_i^{(l)})$. Therefore, a characterization of these marginal distributions is essential for further analytical progress.
The marginal distributions $G_i^{(l)}(\hat{\omega}^{(l)})$ do not have, in general, a simple analytical expression.
We will derive the expression for them when the link frequencies are sampled from a Gaussian distribution with mean $\Omega_1^{(l)}/N_0$, and standard deviation $1/(\tau^{(l)}_1\sqrt{N_0-1})$ ($\tau^{(l)}_1$ is a real number). The distributions $G_i^{(l)}(\hat{\omega}_i^{(l)})$ of $\hat{\Omega}_i^{(l)}$ are Gaussian distribution with means $\langle \hat{\omega}_i^{(l)}\rangle=(\sum_{j<i}a^{(l)}_{ij}-\sum_{j>i}a^{(l)}_{ij})\Omega_1/N_0$ and the covariance matrices $\textbf{C}^{(l)}$ given by $C^{(l)}_{ij}=\langle \hat{\omega}_i^{(l)} \hat{\omega}_j^{(l)} \rangle_c=\frac{[L^{(l)}_{[0]}]_{ij}}{(N_0-1)(\tau^{(l)}_1)^2}$

This covariance matrix $\textbf{C}^{(l)}$ has $(N_0 - 1)$ eigenvalues given by \( \lambda = 1/(\tau^{(l)}_1)^2 \) and one zero eigenvalue \( \lambda = 0 \) corresponding to the eigenvector

\[
\frac{{\mathbf{1}}_{N_0}}{\sqrt{N_0}} = \left(1, 1, \dots, 1\right)/\sqrt{N_0}.
\]

This means that we always have the constraint

\[
\sum_{i=1}^{N_0} \left[ \hat{\omega}_i^{(l)} - \langle \hat{\omega}_i^{(l)} \rangle \right] \sqrt{N_0} = 0.
\]

We can introduce this constraint as a delta function in the expression for the joint distribution \( \hat{G}^{(l)}(\hat{\omega}^{(l)}) \) of the vector \( \hat{\omega}^{(l)} \). Given the zero eigenvalue in the covariance matrix, we write the joint Gaussian distribution \( \hat{G}^{(l)}(\hat{\omega}^{(l)}) \) as,

\[
\hat{G}^{(l)}(\hat{\omega}^{(l)}) = C e^{-F(\hat{\omega}^{(l)})} \, \delta \left( \sum_{i=1}^{N_0} \left[ \hat{\omega}_i^{(l)} - \langle \hat{\omega}_i^{(l)} \rangle \right] \sqrt{N_0} \right),
\]

where \( \delta(x) \) denotes the Dirac delta function, and

\[
F(\hat{\omega}^{(l)}) = \frac{(\tau_1^{(l)})^2}{2} \sum_{i=1}^{N_0} \left( \hat{\omega}_i^{(l)} - \langle \hat{\omega}_i^{(l)} \rangle \right)^2, \quad 
C = \left( \frac{\tau^{(l)}_1}{\sqrt{2\pi}} \right)^{N_0-1}. 
\]

The marginal probability \( G_i^{(l)}(\hat{\omega}^{(l)}) \) is given by

\[
G_i^{(l)}(\hat{\omega}^{(l)}) = \int \prod_{n \ne i} d\hat{\omega}_n^{(l)} \, \hat{G}^{(l)}(\hat{\omega}^{(l)}).
\]

By expressing the delta function in its integral form:

\[
\delta(x) = \frac{1}{2\pi} \int_{-\infty}^{\infty} dz \, e^{izx}, 
\]

we get the final expression for the marginal distribution \(G_i^{(l)}(\hat{\omega}_i^{(l)}) \). By setting \( c = \frac{N_0}{N_0-1} \), we obtain:
\begin{widetext}
\[
G_i^{(l)}(\hat{\omega}_i^{(l)}) = \frac{C}{2\pi} \int dz \int \prod_{n \ne i} d\hat{\omega}_n^{(l)} \, 
e^{-F(\hat{\omega}_i^{(l)})} \exp\left( \text{i}z \sum_{n=1}^{N_0} \left[ \hat{\omega}_n^{(l)} - \langle \hat{\omega}_n^{(l)} \rangle \right] \sqrt{N_0} \right).
\]
\end{widetext}
This evaluates to,
\begin{widetext}
\[
G_i^{(l)}(\hat{\omega}^{(l)}_i) = \frac{1}{2\pi} \int dz \, \exp\left( -\frac{(\tau^{(l)}_1)^2}{2} \left[ \hat{\omega}_i^{(l)} - \langle \hat{\omega}_i^{(l)} \rangle \right]^2 - \frac{z^2}{2(\tau^{(l)}_1)^2 c} + iz \left[ \hat{\omega}_i^{(l)} - \langle \hat{\omega}_i^{(l)} \rangle \right] \sqrt{N_0} \right),
\]
\end{widetext}

and implies that,

\[
G_i^{(l)}(\hat{\omega}_i^{(l)}) = \frac{\tau_1^{(l)}}{ \sqrt{\frac{2\pi}{c}}} \exp\left( -\frac{(\tau^{(l)}_1)^2 c}{2} \left( \hat{\omega}_i^{(l)} - \langle \hat{\omega}_i^{(l)} \rangle \right)^2 \right).
\]

Hence, we have finally derived the expression for the marginal distribution of the projections of the link signals on the nodes in each layer. Now, by considering the Gaussian distribution $\mathcal{N}(\Omega_0^{(l)},1/\tau_0^{(l)})$ for the node frequencies of the layers, we can solve the integrals in the Eqs. (\ref{node_ann}, \ref{link_down_ann}). Then Eq. \eqref{node_ann} gives
\begin{equation}
\label{m2_bw_1}
    \sigma \tau_0^{(l)} R_0^{(l')}R_-^{(l')}R_-^{(l)}\sqrt{\pi/2}\frac{e^{-A/4}}{2}[I_0(A/4)+I_1(A/4)]=1,
\end{equation}
and the Eq. \eqref{link_down_ann} implies that
\begin{equation}
\label{m2_bw_2}
 R_-^{(l)}=\sigma \tau_1^{(l)} R_0^{(l')}R_-^{(l')}\sqrt{c\pi/2}\frac{e^{-B/4}}{2}[I_0(B/4)+I_1(B/4), 
\end{equation}
where $A=(\sigma \tau_0^{(l)}R_0^{(1)}R_0^{(2)}R_-^{(1)}R_-^{(2)})^2$, \\$B=c(\sigma \tau_1^{(l)}R_0^{(l')}R_-^{(l')})^2$, and $I_0,I_1$ are the modified Bessel functions of the first kind of order zero and one, respectively. These two Eqs. (\ref{m2_bw_1}, \ref{m2_bw_2}) fully describe the backward transition scenario of the globally coupled multilayer network model M2. 

\subsection{Verification of theoretical derivations in globally coupled scenario}
\begin{figure}[!ht]
    \centering
    \includegraphics[width=1.1\linewidth]{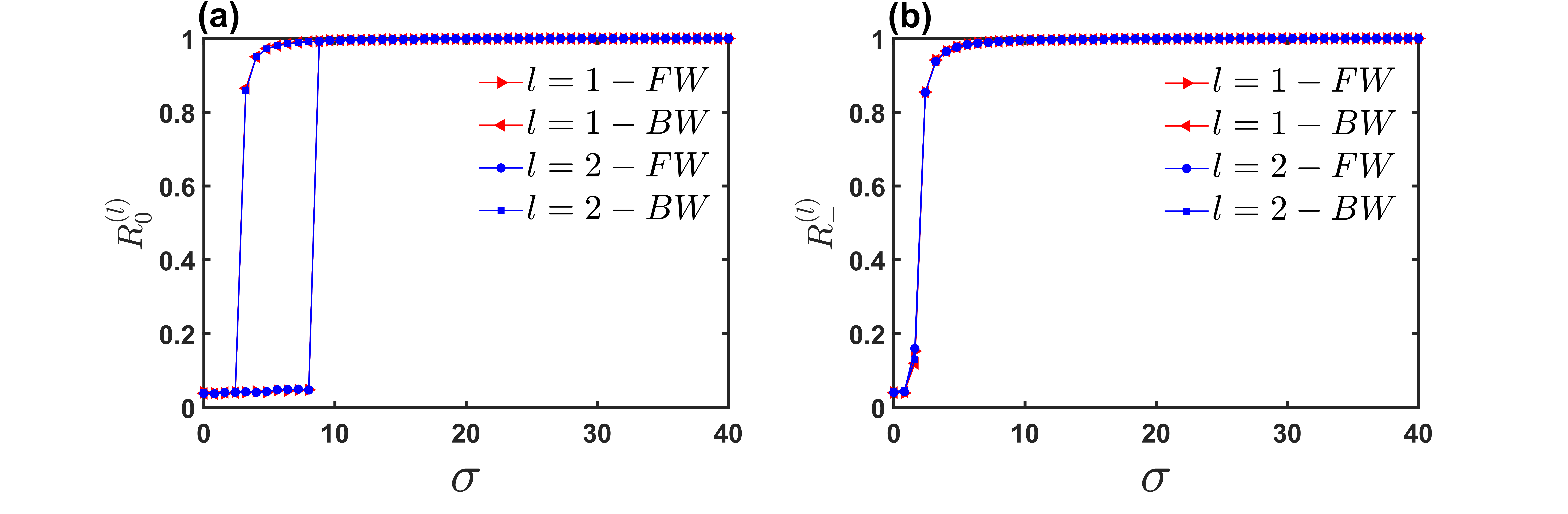}
    \caption{\textbf{The backward and forward explosive phase transition profiles of signals in the multilayer network M1.} Panel (a) displays the synchronization transition of the node signals of the two layers, and panel (b) shows the explosive phase transition of the down-link signal. In this multilayer framework M1, both layers are globally coupled networks of $N_0=500$ nodes.}
    \label{ata_m1}
\end{figure}

To verify the obtained analytical expression, we first numerically simulate the backward and forward phase transition profile of the amplitude of the order parameters of the node (see Fig. \ref{ata_m1}(a)) and down-link (Fig. \ref{ata_m1}(b)) signals of the two layers. 

They show the same for multilayer network M1, and this behavior holds true even for the remaining models M2 and M3. Therefore, we consider $R_0^{(l)}=R_0^{(l')}$ and $R_-^{(l)}=R_-^{(l')}$ in our last two Eqs. (\ref{m2_bw_1}, \ref{m2_bw_2}); consequently, we have the final form of these equations as,
\begin{equation}
    \sigma \tau_0^{(l)} R_0^{(l)}(R_-^{(l)})^2\sqrt{\pi/2}\frac{e^{-A/4}}{2}[I_0(A/4)+I_1(A/4)]=1,
    \label{ana1_m2}
\end{equation}
\begin{equation}
 \sigma \tau_1^{(l)} R_0^{(l)}\sqrt{c\pi/2}\frac{e^{-B/4}}{2}[I_0(B/4)+I_1(B/4) =1
 \label{ana2_m2}
\end{equation}
where $A=(\sigma \tau_0^{(l)}(R_0^{(l)}R_-^{(l)})^2)^2$ and $B=c(\sigma \tau_1^{(l)}R_0^{(l)}R_-^{(l)})^2$.
\par Following the same derivations given above, we can have the relations between the amplitude of the order parameters of the nodes and down-links of the layers for the globally coupled multilayer network M1 as
\begin{equation}
    \sigma \tau_0^{(l)} R_0^{(l)}(R_-^{(l)})^2\sqrt{\pi/2}\frac{e^{-A/4}}{2}[I_0(A/4)+I_1(A/4)]=1,
    \label{ana1_m1}
\end{equation}
\begin{equation}
 \sigma \tau_1^{(l)} \sqrt{c\pi/2}\frac{e^{-B/4}}{2}[I_0(B/4)+I_1(B/4)=1,
 \label{ana2_m1}
\end{equation}
where $A=(\sigma \tau_0^{(l)}(R_0^{(l)}R_-^{(l)})^2)^2$ and $B=c(\sigma \tau_1^{(l)}R_-^{(l)})^2$.
\par
\begin{figure*}[t]
     \centering
     \includegraphics[width=\textwidth]{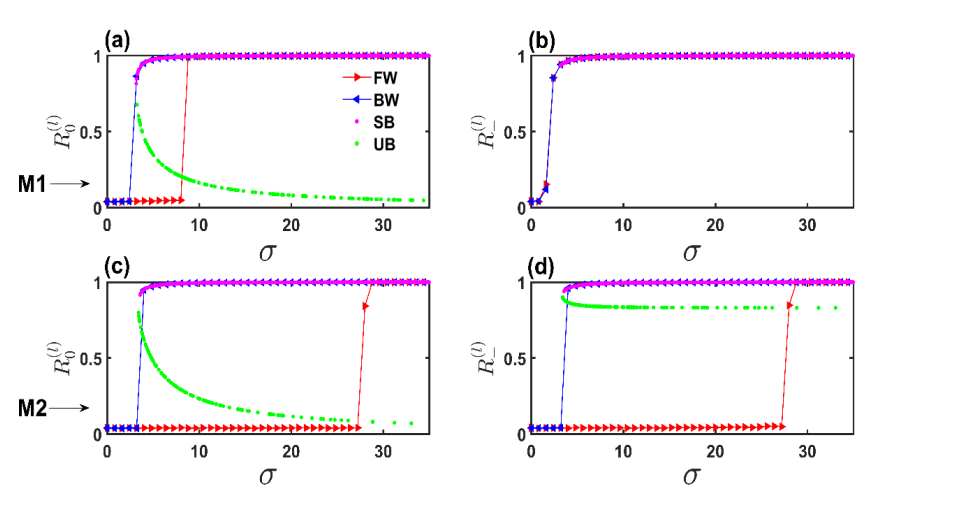}
     \caption{{\bf The backward and forward discontinuous phase transition of the amplitude of the order parameters.} For the multilayer network, in which both layers are globally coupled, (a), (c) are the phase transition profiles of nodes, and (b), (d) are the phase transition profiles of the down-links within each layer. The first row and second row are for the multilayer networks M1, and M2, respectively. The dotted curves in magenta and green are the stable branch (SB) and unstable branch (UB), respectively, predicted from the analytical derivations. The red and blue curves are the forward, and backward transitions, respectively. We take $N_0=500$ nodes in each layer, and for the Gaussian frequency distributions of the nodes and links, we consider $\Omega_0^{(l)}=\Omega_1^{(l)}=0$, $\tau_0^{(l)}=\tau_1^{(l)}=1$ values for the layers $l=1,2$.}
     \label{ata_m1_m2}
 \end{figure*}

\begin{figure}[!ht]
     \centering
     \includegraphics[width=1.1\linewidth]{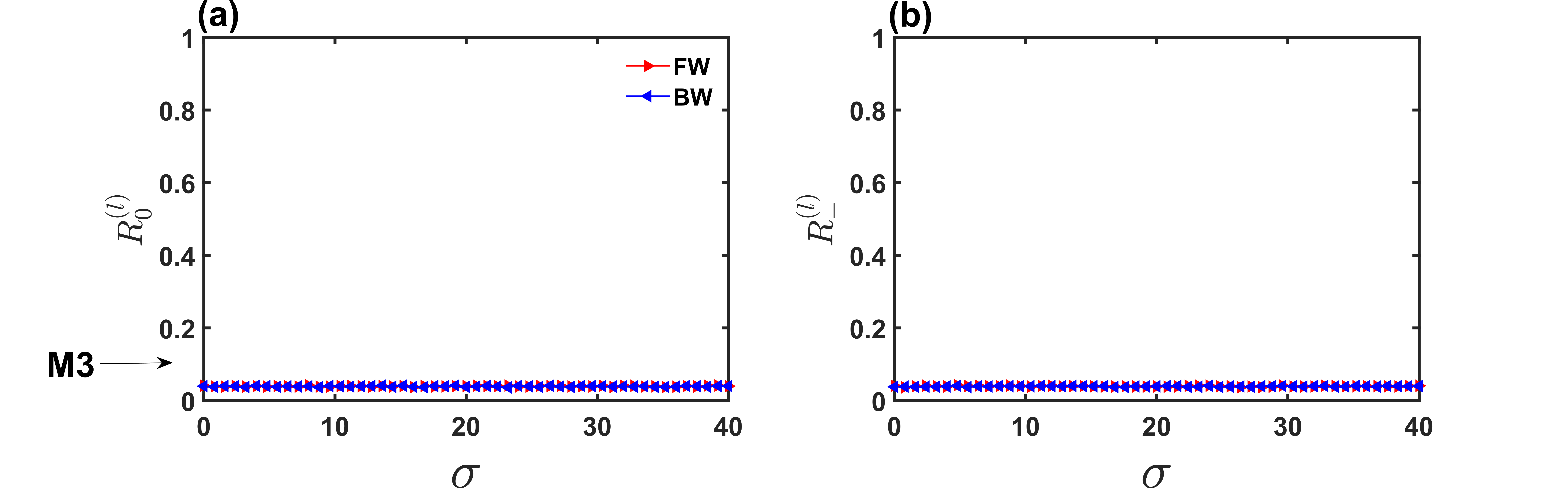}
     \caption{\textbf{The backward and forward profiles of signals in the multilayer network M3.} Panels (a), and (b), display, respectively, to the amplitude of the order parameters of the node signals and down-link signals. In this framework M3, the layer are globally coupled networks of $500$-nodes.}
     \label{ata_m3}
 \end{figure}

 The verification of the analytical derivations given in Eqs. (\ref{ana1_m1}, \ref{ana2_m1}) for the model M1 and Eqs. (\ref{ana1_m2}, \ref{ana2_m2}) for the model M2, with the numerical simulations of the nodes (given by Eq. \eqref{eq_gen_node}) and links (given by Eq. \eqref{eq_gen_link1}) dynamics in a globally coupled scenario, is presented in Fig. \ref{ata_m1_m2}. The phase transition profiles (both backward and forward) of nodes and links of the multilayer model M1 are given in Figs. \ref{ata_m1_m2}(a), and \ref{ata_m1_m2}(b), respectively. We observe a hysteresis in the nodes' profile, but there is no hysteresis in the links' profile. The critical value of the coupling strength for the backward transition of nodes and down-link remains the same in the multilayer models M1 and M2. The critical values of the forward transitions of the nodes and the links for these two models are not the same. The critical value of the model M2 gets significantly increased. Therefore, we observe a broad bistability region in each of the phase transition profiles of node and down-link in the model M2 (see Fig. \ref{ata_m1_m2}(c,d)). However, the phases of the nodes and down-links in the globally coupled multilayer model M3 are not synchronized (see Fig. \ref{ata_m3}). 
 \newpage
 \section{Conclusion}

In this work, we investigated the synchronization of topological signals on multilayer simplicial complexes with adaptive interlayer coupling. Focusing on bilayer systems composed of two-dimensional simplicial complexes, we studied the collective dynamics of node, link, and triangle signals under both same-dimensional and cross-dimensional coupling schemes. By combining numerical simulations with analytical results obtained in the annealed and fully connected limits, we showed how adaptive feedback between layers reshapes the synchronization transitions of higher-order signals.

\par Our results reveal that adaptive coupling involving projections of link signals onto adjacent simplicial spaces can substantially delay the onset of synchronization. In particular, coupling to node and up-link signals acts as an effective obstruction mechanism, shifting the critical coupling strength to significantly larger values compared to non-adaptive or purely same-dimensional settings. This effect is robust across different network geometries, including configuration models and network geometries with flavor.

\par By exploiting the Hodge decomposition of link signals, we clarified the distinct roles played by irrotational, solenoidal, and harmonic components in the synchronization process. The projected down-link and up-link signals selectively probe these components and provide a natural geometric interpretation of the observed transitions. Our analysis further demonstrates that, despite the increased complexity introduced by multilayer and adaptive interactions, analytical progress remains possible and accurately captures the numerical behavior in appropriate limits.

\par These findings contribute to the growing understanding of dynamical processes on higher-order and multilayer networks, highlighting how adaptive and cross-dimensional interactions can qualitatively alter collective behavior. Beyond synchronization, the framework developed here can be extended to study other dynamical phenomena on multilayer simplicial complexes, including pattern formation, control, and information processing in systems with higher-order interactions.
\section{Methods}
In our study, we use tools from algebraic topology to model and analyze a higher-order multilayer framework. A higher-order multilayer network can be formed by a set of an arbitrary $L$ number of oriented simplicial complexes, which gives the underlying intralayer topology structure of each of the $L$ layers, and they are connected by some simplices, which give the interlayer architecture of the multilayer network. In this section, we begin by introducing the notion of a simplicial complex, chain, co-chain, and the associated boundary operators. Then we define each of them separately for each layer of the higher-order multilayer network.\\

\subsection{Oriented Simplicial Complexes}
\textbf{Simplices:}
A $d$-simplex $\sigma^d$ is the convex hull of $d+1$ affinely independent points
$i_0, i_1, \dots, i_d$ in a Euclidean space of dimension greater than or equal to $d$. It is denoted by
\[
\sigma^d = [i_0, i_1, \dots, i_d],
\]
where the vertices $i_0, i_1, \dots, i_d$ are distinct. The dimension of the simplex
is $d$.\\ \textbf{Faces of a simplex:}
Any simplex obtained by taking a subset of the vertices of a simplex $\sigma^d$
defines a face of $\sigma^d$. In particular, a $k$-face
($0 \le k \le d$) of $\sigma^d = [i_0, i_1, \dots, i_d]$ is a $k$-simplex of the form
\[
[i_{j_0}, i_{j_1}, \dots, i_{j_k}],
\]
where $\{j_0, j_1, \dots, j_k\} \subset \{i_0,i_1,\dots,i_d\}$.\\
\textbf{Simplicial Complex:} A $D$-dimensional simplicial complex $\mathcal{K}$ is a collection of simplices
of dimension at most $D$ satisfying the following properties:
\begin{itemize}
  \item There exists at least one simplex of dimension $D$ in the collection $\mathcal{K}$.
  \item If a simplex $\sigma \in \mathcal{K}$, then all faces of $\sigma$ also
  belong to $\mathcal{K}$.
  \item The intersection of any two simplices in $\mathcal{K}$ is either empty
  or a common face of both.
\end{itemize}
\textbf{Oriented simplicial complex:}
A $D$-dimensional oriented simplicial complex $\mathcal{K}$ is a simplicial
complex in which each simplex is assigned an orientation.
For example, in a 3-dimensional space,
\begin{itemize}
  \item $\sigma^0$: An oriented $0$-simplex is a vertex with a trivial or no orientation.
  \item $\sigma^1$: An oriented $1$-simplex $[i,j]$ represents a directed edge from vertex $i$ to vertex $j$.
  \item $\sigma^2$: An oriented $2$-simplex $[i,j,k]$ represents an oriented filled triangle with the vertices $i,j,k$,  whose orientation is determined by the cyclic ordering of the
vertices. Reversing the ordering corresponds to reversing the orientation
(e.g., clockwise versus anticlockwise).
  \item $\sigma^3$: An oriented $3$-simplex $[i,j,k,l]$ represents an oriented tetrahedron with vertices $i,j,k,l$. Its orientation is determined by the ordering of the
vertices and can be interpreted as the sign of the associated volume form;
reversing the ordering corresponds to reversing the orientation.
\end{itemize}
In our model, we consider a multilayer structure of $L$ layers comprising $L$ oriented simplicial complexes, denoted  $\mathcal{K}^{(l)}$ for $l = 1, 2, \dots, L$, each of dimension $D$. These layers are adaptively connected by modulating the coupling strength of the topological signals and their projections through their order parameters. The $l$-th layer has $N_d^{(l)}$ $d$-simplices, and we define the set of $d$-simplices in the $l$-th layer by
\[
\mathcal{S}^{(l)}_d = \left\{ \sigma^d_i \ \middle| \ i = 1, 2, \dots, N_d^{(l)} \right\}.
\]
where $N_d^{(l)}$ is the total number of $d$-simplices in the simplicial complex $\mathcal{K}^{(l)}$.
\subsection{Chain Groups}
The \textbf{$d$-chain group} $C^{(l)}_d(\mathcal{K}^{(l)})$ over the simplicial complex $\mathcal{K}^{(l)}$  is defined as the free abelian group with its basis as the set of all $d$-dimensional simplices in $\mathcal{K}^{(l)}$, i.e.,
\[
C_d^{(l)}(\mathcal{K}^{(l)}) = \left\{ \sum_i a_i \sigma^d_i \,\middle|\, a_i \in \mathbb{Z},\, \sigma^d_i \in \mathcal{S}_d^{(l)} \right\},
\]
where $\mathbb{Z}$ is the set of integer numbers.

Each $d$-chain is a formal linear combination of oriented $d$-simplices with integer coefficients. If we use real coefficients, i.e., \[
C_d^{(l)}(\mathcal{K}^{(l)}; \mathbb{R}) = \text{span}_{\mathbb{R}}(\mathcal{S}_d^{(l)}),
\] then it becomes a real vector space.

\subsection{Cochains/Topological signals}

In a multilayer structure consisting of $L$ simplicial complexes $\{ \mathcal{K}^{(1)}, \mathcal{K}^{(2)}, \dots, \mathcal{K}^{(L)} \}$, where each $\mathcal{K}^{(l)}$ is a $D$-dimensional simplicial complex, we define cochains separately on each layer.

A \textbf{$d$-cochain} ($0\le d \le D$) on a layer $l$ is an operator
\[
\Phi_d^{(l)} : C_d^{(l)} \to \mathbb{R}^n,~n\in \mathbb{N}
\]
that assigns a $n$-dimensional Euclidean vector to each $d$-dimensional simplex in the layer $\mathcal{K}^{(l)}$.

Since $\mathcal{S}_d^{(l)} = \{ \sigma^d_1, \sigma^d_2, \dots, \sigma^d_{N_d^{(l)}} \}$ is the ordered set of basis $d$-simplices of the vector space $C_d^{(l)}$, therefore the cochain $\Phi_d^{(l)}$ can be represented by the vectors $\Phi_d^{(l)}(\sigma^d_i)$ for $i=1,2,\dots,N_d$. Hence, $\Phi_d^{(l)}$ is equivalent to the column vector \[\begin{bmatrix}
\Phi_d^{(l)}(\sigma^d_1) \\
\Phi_d^{(l)}(\sigma^d_2) \\
\vdots \\
\Phi_d^{(l)}(\sigma^d_{N_d^{(l)}})
\end{bmatrix}
\in \mathbb{R}^{nN_d^{(l)}}.
\]

These $d$-cochains are the topological signals defined on each $d$-simplex in the simplicial complex, which are used in defining coboundary operators within each layer independently, unless inter-layer operators are introduced. In our context of the Kuramoto oscillator, the dimension $n$ of the codomain space of all the cochains is taken to be $1$, which represents the phase variable associated with each simplex. Since we focus on phase oscillators throughout the manuscript, we set $n=1$.

\subsection{Boundary Operators}

The topological structure of a simplicial complex is encoded through the boundary operators, which map simplices of dimension $d$ to their $(d-1)$-dimensional faces. For each layer $l$, the $d$-th boundary operator is denoted by
\[
\partial_d^{(l)} : C_d^{(l)} \to C_{d-1}^{(l)},
\]
where $C_d^{(l)}$ and $C_{d-1}^{(l)}$ are the chain groups generated by the $d$-simplices and $(d-1)$-simplices of layer $l$, respectively.

In matrix form, the boundary operator $\partial_d^{(l)}$ can be represented by a matrix $B_d^{(l)} \in \mathbb{R}^{N_{d-1}^{(l)} \times N_d^{(l)}}$, where
\[
[B_d^{(l)}]_{ij} = 
\begin{cases}
\pm 1, & \text{if } (d-1)\text{-simplex } \sigma^{d-1}_i \text{ is a face of } \sigma^d_j, \\
0, & \text{otherwise}.
\end{cases}
\]

The sign $\pm 1$ depends on the relative orientation of the simplices. These matrices satisfy the topological identity:
\[
\partial_{d-1}^{(l)} \circ \partial_d^{(l)} = 0, \quad \text{or equivalently,} \quad B_{d-1}^{(l)} B_d^{(l)} = 0,
\]
which ensures that the boundary of a boundary is always zero—a foundational property in algebraic topology.

These boundary matrices are the backbone of topological computations, such as determining homology groups or defining higher-order Laplacians used in the proposed multilayer Kuramoto model.
\subsection{Hodge decomposition of 1-cochain/link signals} \label{hodge}
 In the analysis of link-signals dynamical processes on simplicial complexes, it is often useful 
to decompose the space of edge signals into orthogonal subspaces capturing distinct structural modes. 
The Hodge decomposition provides exactly such a framework: it splits each edge signal uniquely 
into harmonic, solenoidal, and irrotational components \cite{barbarossa2020topological}. 
This decomposition is intimately connected to the 1-Hodge Laplacian.

For each layer $l$, the edge (link) signal space admits the Hodge decomposition \begin{equation}
\mathbb{R}^{N_1^{(l)}} = \ker\big(L^{(l)}_1\big) 
\oplus \operatorname{Im}\big([B^{(l)}_1]^T\big) 
\oplus \operatorname{Im}\big(B^{(l)}_2\big),
\end{equation}
where $\ker\big(.\big)$, and $\operatorname{Im}\big(.\big)$ represent the kernel and image of a matrix, respectively. Also, the $1$-Hodge Laplacian decomposes into its up- and down-parts as
\begin{equation}
L^{(l)}_1 = [L^{(l)}_1]^{\mathrm{up}} + [L^{(l)}_1]^{\mathrm{down}}
= B^{(l)}_2[B^{(l)}_2]^T + [B^{(l)}_1]^T B^{(l)}_1.
\end{equation}
Accordingly, any link-phase vector $\Phi^{(l)}$ can be written uniquely as the sum of a harmonic, 
a solenoidal, and an irrotational component:
\begin{equation}
\Phi^{(l)} = H^{(l)} + X^{(l)}_1 + X^{(l)}_2,
\end{equation}
with $H^{(l)} \in \ker(L^{(l)}_1)$, $X^{(l)}_1 \in \operatorname{Im}(B^{(l)}_2)$ (so $B^{(l)}_1 X^{(l)}_1=0$) 
and $X^{(l)}_2 \in \operatorname{Im}([B^{(l)}_1]^T)$ (so $[B^{(l)}_2]^T X^{(l)}_2=0$). 
Because $L^{(l)}_1$ annihilates harmonic fields, applying the Laplacian yields only the up- and down- contributions:
\begin{equation}
\begin{aligned}
L^{(l)}_1\Phi^{(l)} &= L^{(l)}_1 X^{(l)}_1 + L^{(l)}_1 X^{(l)}_2 \\
&= [L^{(l)}_1]^{\mathrm{up}} X^{(l)}_1 + [L^{(l)}_1]^{\mathrm{down}} X^{(l)}_2 \\
&= Y_1^{(l)} + Y_2^{(l)},
\end{aligned}
\end{equation}
where $Y_1^{(l)}$ and $Y_2^{(l)}$ lie in the range of $[L^{(l)}_1]^{\mathrm{up}}$ and $[L^{(l)}_1]^{\mathrm{down}}$, respectively. That means $Y_1^{(l)}=[L^{(l)}_1]^{\mathrm{up}}\Phi^{(l)}$ and $Y_2^{(l)}=[L^{(l)}_1]^{\mathrm{down}}\Phi^{(l)}$.
\subsection{Up-link and down-link signals: projections of link signal's dynamics}
\label{uplink-downlink}
To connect the Hodge decomposition with the dynamics, we define the down-link and up-link signals as projections of the link-phase vector onto adjacent simplicial spaces. For each layer $l$, the down-link signal is obtained by projecting link phases onto nodes,
\begin{equation}
\Phi^{(l)}_{-} = B^{(l)}_1 \Phi^{(l)},
\end{equation}
while the up-link signal is obtained by projecting link phases onto triangles,
\begin{equation}
\Phi^{(l)}_{+} = [B^{(l)}_2]^T \Phi^{(l)}.
\end{equation}
The harmonic component does not contribute to either projection, as it lies in the kernel of both operators. These projections isolate complementary dynamical modes of the link signal and form the basis of the adaptive couplings introduced in the multilayer models.\\
\section*{Acknowledgements}
P.K.P acknowledges financial support from the University Grants Commission (UGC), India, through the Senior Research Fellowship (SRF) scheme [NTA Ref. No. 191620041800].
\section*{Author Contributions}
P.K.P. and D.G. conceived the study and designed the research methodology. P.K.P. developed the computational framework, performed the numerical simulations, analyzed the results, and wrote the first draft of the manuscript. J.K. and D.G. contributed to the conceptual development of the work, supervised the research, assisted in the interpretation and visualization of the results, and coordinated the project. All authors discussed the results, reviewed and edited the manuscript, and approved the final version of the paper.
\section*{Competing Interests}
The authors declare no competing interests.
\section*{Data Availability}
All data needed to evaluate the findings of the paper are available within the
paper itself. Additional data related to this paper are available from the
corresponding author upon reasonable request.\\
\section*{Code Availability}
The codes used in the simulations for this article are available openly on the
GitHub repository \cite{Pal2026Code}.


%

\end{document}